\documentclass[12pt]{iopart}
\usepackage{graphicx}
\usepackage{cite}
\usepackage{amsfonts}
\usepackage{amsthm}

\newcommand{\id}{\mathbf{1}}
\newcommand{\bra}[1]{\langle #1 \hspace{0mm} |}
\newcommand{\ket}[1]{|\hspace{0mm} #1 \rangle}
\newcommand{\braket}[2]{\langle #1 | #2 \rangle}
\newcommand{\tensormin}{\otimes_\mathrm{min}}
\newcommand{\tensormax}{\otimes_\mathrm{max}}
\newcommand{\gtensormax}{\overline{\otimes}_\mathrm{max}}

\begin{document}

\topical[Generalized Probability Theories]
      {Generalized Probability Theories:\\What determines the structure of quantum theory?}
\author{Peter Janotta and Haye Hinrichsen}
\address{Universit\"at W\"urzburg, Fakult\"at f\"ur Physik und Astronomie,  97074 W\"urzburg, Germany}

\begin{abstract}
The framework of generalized probabilistic theories is a powerful tool for studying the foundations of quantum physics. It provides the basis for a variety of recent findings that significantly improve our understanding of the rich physical structure of quantum theory. This review paper tries to present the framework and recent results to a broader readership in an accessible manner.     
To achieve this, we follow a constructive approach. Starting from few basic physically motivated assumptions we show how a given set of observations can be manifested in an operational theory. Furthermore, we characterize consistency conditions limiting the range of possible extensions. In this framework classical and quantum theory appear as special cases, and the aim is to understand what distinguishes quantum mechanics as the fundamental theory realized in nature. It turns out non-classical features of single systems can equivalently result from higher dimensional classical theories that have been restricted. Entanglement and non-locality, however, are shown to be genuine non-classical features.
\end{abstract}

\parskip 2mm 

\section{Introduction}

Quantum theory is considered to be the most fundamental and most accurate physical theory of today. Although quantum theory is conceptually difficult to understand, its mathematical structure is quite simple. What determines this particularly simple and elegant mathematical structure? In short: Why is quantum theory as it is?

Addressing such questions is the aim of investigating the \textit{foundations of quantum theory}. In the past this field of research was sometimes considered as an academic subject without much practical impact. However, with the emergence of quantum information theory this perception has changed significantly and both fields started to fruitfully influence each other \cite{hardy10, naturereview}. Today fundamental aspects of quantum theory attract increasing attention and the field belongs to the most exciting subjects of theoretical physics.

In this topical review we will be concerned with a particular branch in this field, namely, with so-called \textit{Generalized Probabilistic Theories} (GPTs), which provide a unified theoretical framework in which classical and quantum theory emerge as special cases. Presenting this concept in the language of statistical physics, we hope to establish a bridge between the communities of classical statistical physics and quantum information science. 

The early pioneers of quantum theory were strongly influenced by positivism, a school of philosophy postulating that a physical theory should be built and verified entirely on the basis of accessible sensory experience. Nevertheless the standard formulation of quantum theory involves additional concepts such as global complex phases which are not directly accessible. The GPT framework, which is rooted in the pioneering works by Mackey, Ludwig and Kraus~\cite{mackey63,ludwig83,ludwig85,kraus83}, tries to avoid such concepts as much as possible by defining a theory operationally in terms of preparation procedures and measurements.

As measurement apparatuses yield classical results, GPTs are exclusively concerned with the classical \textit{probabilities} of measurement outcomes for a given preparation procedure. As we will see below, classical probability theory and quantum theory can both be formulated within this unified framework. Surprisingly, starting with a small set of basic physical principles, one can construct a large variety of other consistent theories with different measurement statistics. This generates a whole spectrum of possible theories, in which classical and quantum theory emerge just as two special cases. Most astonishingly, various properties thought to be special for quantum theory turn out to be quite general in this space of theories. As will be discussed below, this includes the phenomenon of entanglement, the no-signaling theorem, and the impossibility to decompose a mixed state into unique ensemble of pure states.

Although GPTs are defined operationally in terms of probabilities for measurement outcomes, it is not immediately obvious how such a theory can be constructed from existing measurement data. In this work we shed some light on the process of building theories in the GPT framework on the basis of a set of given experimental observations, making the attempt to provide step-by-step instructions how a theory can be constructed. 

The present Topical Review is written for readers from various fields who are interested to learn something about the essential concepts of GPTs. Our aim is to explain these concepts in terms of simple examples, avoiding mathematical details whenever it is possible. We present the subject from the perspective of model building, making the attempt to provide step-by-step instructions how a theory can be constructed on the basis of a given set of experimental observations. To this end we start in Sect. \ref{sec:datatable} with a data table that contains all the available statistical information of measurement outcomes. In Sect.~\ref{sec:statesandeffects} the full space of possible experimental settings is then grouped into equivalence classes of observations, reducing the size of the table and leading to a simple prototype model. As shown in Sect.~\ref{sec:linear} this prototype model has to be extended in order to reflect possible deficiencies of preparations and measurements, leading in turn to new suitable representations of the theory. This extension can be chosen freely within a certain range limited by certain consistency conditions (see Sect.~\ref{sec:consistency}). Depending on this choice the extended theory finally allows one to make new predictions in situations that have not been examined so far. Within this framework we discuss three important minimal systems, namely, the classical \textit{bit}, then quantum bit (\textit{qubit}), as well as the so-called \textit{gbit}, which can be thought of as living somewhere between classical and quantum theory.

In Sect.~\ref{sec:nonclassicalbyrestriction} we devote our attention to the fact that any non-classical system is equivalent to a classical system in a higher-dimensional state space combined with certain constraints. However, this equivalence is only valid as long as non-composite (single) systems are considered. Turning to bipartite and multipartite systems the theory has to be complemented by a set of composition rules in the form of a suitable tensor product (see Sect. \ref{sec:jointsystems}). Again it turns out that there is some freedom in choosing the tensor product, which determines the structure of a GPT to a large extent. Finally, in Sect. \ref{sec:nonlocalcorrelations} we discuss nonlocal correlations as a practical concept that can be used to experimentally prove the existence of non-classical entanglement in composite systems without the need to rely on a particular theory.

For beginners it is often difficult to understand the construction of a non-classical theory without introducing concepts such as Hilbert spaces and state vectors. For this reason we demonstrate how ordinary quantum mechanics fits into the GPT framework, both for single systems in Sect. \ref{sec:QTsinglefeatures} and for composite systems in Sect. \ref{sec:QTjointfeatures}.

\section{Generalized probabilistic theories: Single systems}

\subsection{Preparation procedures and measurements}
\label{sec:datatable}

\begin{figure}[t]
\centering\includegraphics[width=120mm]{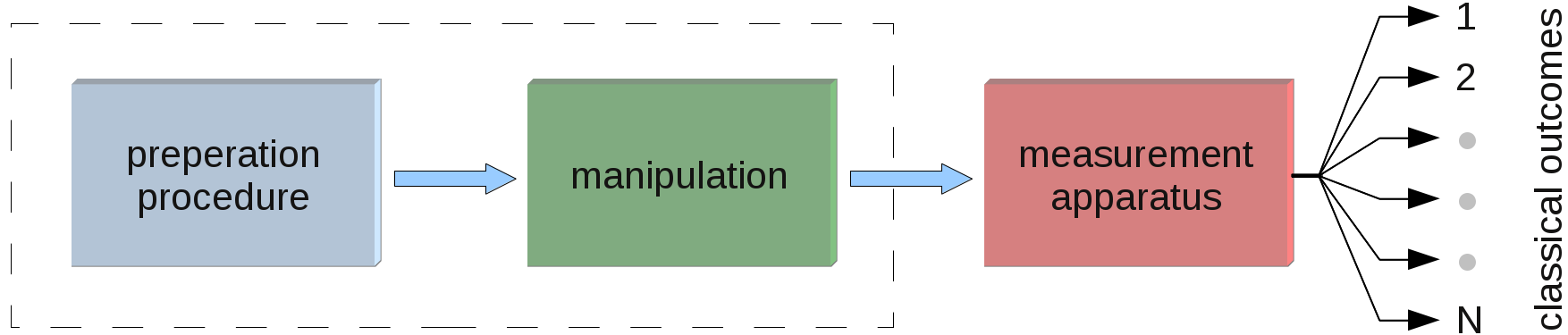}
\caption{\small Typical experimental setup consisting of a preparation procedure, a sequence of intermediate manipulations, and a final measurement with a certain set of possible classical outcomes (see text). The intermediate manipulations can be thought of as being part of either the preparation procedure (dashed box) or the measurement. }
\label{fig:expsetup}
\end{figure}

As sketched schematically in Fig.~\ref{fig:expsetup}, a typical experimental setup in physics consists of a preparation procedure, possibly followed by a sequence of manipulations or transformations, and a final measurement. For example, a particle accelerator produces particles in a certain physical state which are then manipulated in a collision and finally measured by detectors. Since the intermediate manipulations can be thought of as being part of either the preparation procedure or the measurement, the setup can be further abstracted to preparations and measurements only\footnote{In standard quantum theory the absorption of intermediate transformations into the preparation procedure corresponds to the Schr\"odinger picture, the absorption into the measurement to the Heisenberg picture.}. 

We can think of a measurement apparatus as a physical device which is prepared in a spring-loaded non-equilibrium idle state. During the measurement process the interaction of the physical system with the device releases a cascade of secondary interactions, amplifying the interaction and eventually leading to a \textit{classical} response that can be read off by the experimentalist. This could be, for example, an audible 'click' of a Geiger counter or the displayed value of a voltmeter.

In practice a measurement device produces either digital or analog results. For analog devices there are in principle infinitely many possible outcomes, but due to the final resolution the amount of information obtained during the measurement is nevertheless finite. Thus, for the sake of simplicity, we will assume that the number of possible outcomes in a measurement is finite.

For an individual measurement apparatus we may associate with each of the possible outcomes a characteristic one-bit quantity which is '1' if the result occurred and '0' otherwise. In this way a measurement can be decomposed into mutually excluding classical bits, as sketched in Fig.~\ref{fig:expsetup}. Conversely every single measurement can be interpreted as a joint application of such fundamental 1-bit measurements. 

If we are dealing with several different measurement devices the associated classical bits are of course not necessarily mutually excluding. This raises subtle issues about coexistence, joint measurability, mutual disturbance and commutativity~\cite{heinosaari10,reeb13}, the meaning of a '0' if the measurement fails, and the possibility to compose measurement devices out of a given set of 1-bit measurements. For simplicity let us for now neglect these issues and return to some of the points later in the article.

\subsection{Toolbox and probability table}

In practice we have only a limited number of preparation procedures and measurement devices at our disposal. It is therefore meaningful to think of some kind of `toolbox' containing a finite number of 1-bit measurements labeled by $k=1\ldots K$ and a finite number of preparation procedures labeled by $\ell=1\ldots L$. As mentioned before, if the range of preparations and measurements is continuous, we assume for simplicity that the finite accuracy of the devices will essentially lead to the same situation with a finite number of elements. Our aim is to demonstrate how the GPT approach can be used to construct a physical theory on the basis of such a toolbox containing $K$~measurement devices and $L$~preparation methods. 

With each pair of a 1-bit measurement $k$ and a preparation procedure $\ell$ we can set up an experiment which produces an outcome $\chi_{k \ell}\in\{0,1\}$. An important basic assumption of the GPT framework is that experiments can be repeated under identical conditions in such a way that the outcomes are statistically independent. Repeating the experiment the specific outcome $\chi_{k \ell}$ is usually not reproducible, instead one can only reproducibly estimate the probability 
\begin{equation}
 p_{k \ell} = \langle \chi_{k \ell} \rangle
\end{equation}
to obtain the result $\chi_{k\ell}=1$ in the limit of infinitely many experiments. For a given toolbox the values of $p_{k\ell}$ can be listed in a probability table. This data table itself can already be seen as a precursor of a physical model. However, it just reproduces the observable statistics and apart from the known probabilities it has no predictive power at all. Moreover, the table may grow as we add more preparation and measurement devices. In order to arrive at a meaningful physical theory, we thus have to implement two important steps, namely,
\begin{enumerate}
 \item to remove all possible redundancies in the probability table, and
 \item to make reasonable assumptions which allow us to predict the behavior of elements which are not yet part of our toolbox.
\end{enumerate}

\subsection{Operational equivalence, states and effects}
%
\label{sec:statesandeffects}

In order to remove redundancies in the probability table let us first introduce the notion of operational equivalence. Two preparation procedures are called \textit{operationally equivalent} if it is impossible to distinguish them experimentally, meaning that any of the available measurement devices responds to both of them with the same probability. Likewise two one-bit measurements are called operationally equivalent if they both respond with the same probability to any of the available preparation procedures.

The notion of operational equivalence allows one to define equivalence classes of preparations and one-bit measurements. Following the terminology introduced by Ludwig and Kraus~\cite{ludwig83,kraus83} we will denote these classes as states and effects:
\begin{itemize}
 \item 
 A \emph{state} $\omega$ is a class of operationally equivalent preparation procedures.
 \item
 An \emph{effect} $e$ is a class of operationally equivalent 1-bit measurements.
\end{itemize}
This allows us to rewrite the probability table in terms of states and effects, which in practice means to eliminate identical rows and columns in the data table. Enumerating effects by $\{e_1, e_2, \ldots, e_M \}$ and states by $\{\omega_1, \omega_2,\ldots, \omega_N \}$ one is led to a reduced table of size $M \times N$, the so-called \textit{fundamental probability table}. 

If we denote by $e(\omega)=p(e|\omega)$ the probability that an experiment chosen from the equivalence classes $e$ and $\omega$ produces a '1', the matrix elements elements of the fundamental probability table can be written as
\begin{equation}
\label{ProbabilityTable}
p_{ij}=\langle \chi_{ij} \rangle = e_i(\omega_j)\,.
\end{equation}
Obviously, this table contains all the experimentally available information. Since effects and states are defined as equivalence classes, it is ensured that no column (and likewise no row) of the table appears twice. 

Note that the later inclusion of additional measurement apparatuses might allow the experimentalist to distinguish preparation procedures which were operationally equivalent before, splitting the equivalence class into smaller ones. This means that a state may split into several states if a new measurement device is added to the toolbox. The same applies to effects when additional preparation procedures are included.

As the introduction of equivalence classes described above eliminates only identical rows and columns, the fundamental probability table can be still very large. In addition, there may be still linear dependencies among rows and columns. As we will see below, these linear dependencies can partly be eliminated, leading to an even more compact representation, but they also play an important role as they define the particular type of the theory.

\subsection{Noisy experiments and probabilistic mixtures of states and effects}
%
Realistic experiments are noisy. This means that a preparation procedure does not always create the physical object in the same way, rather the preparation procedure itself may randomly vary in a certain range. Similarly, a measurement is noisy in the sense that the measurement procedure itself may vary upon repetition, even when the input is identical. In the GPT framework this kind of classical randomness is taken into account by introducing the notion of \textit{mixed} states and effects.

\begin{figure}[t]
\centering\includegraphics[width=135mm]{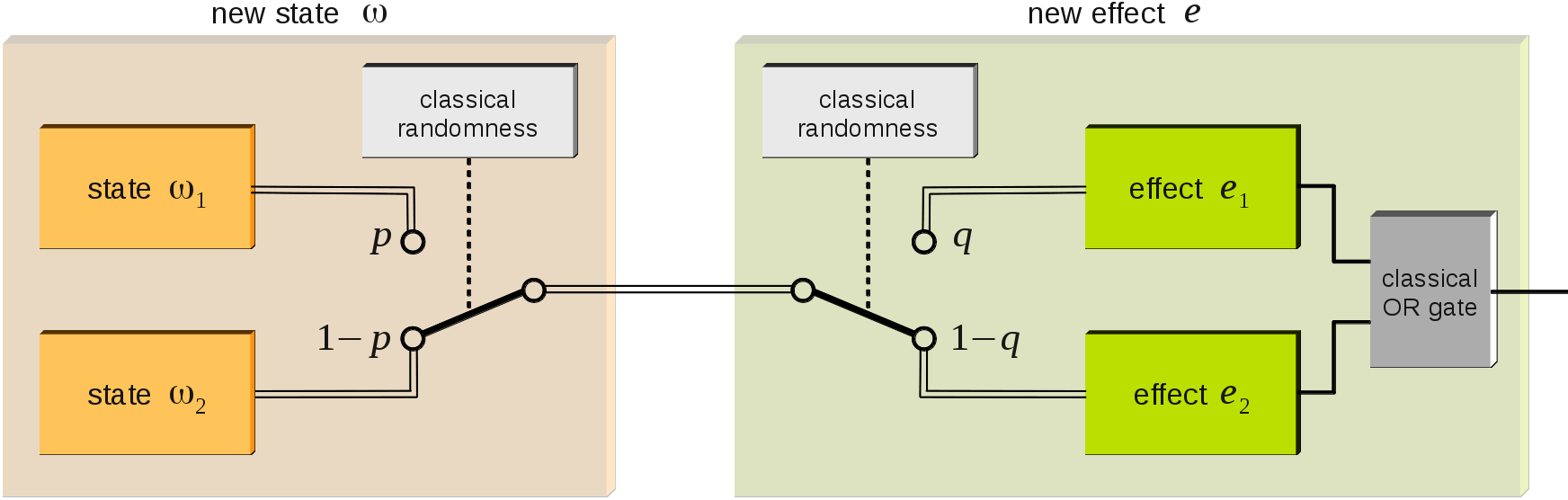}
\caption{New states and effects can be generated by probabilistically mixing the existing ones, illustrated here in a simple example  (see text).}
\label{fig:convex}
\end{figure}

The meaning of probabilistic mixtures is illustrated for the special case of bimodal noise in Fig.~\ref{fig:convex}. On the left side of the figure a classical random number generator selects the preparation procedure $\omega_1$ with probability $p$ and another preparation procedure $\omega_2$ with probability $1-p$. Similarly, on the right side another independent random number generator selects the effect $e_1$ with probability $q$ and the effect $e_2$ otherwise, modeling a noisy measurement device.

If we apply such a noisy measurement to a randomly selected state, all what we get in the end is again a 'click' with certain probability $P$. In the example shown in Fig.~\ref{fig:convex}, this probability is given by
\begin{equation}
\label{example}
\fl\qquad
P = p\,q\,e_1(\omega_1) \,+\, p\,(1-q)\,e_2(\omega_1) + (1-p)\,q\,e_1(\omega_2) \,+\, (1-p)\,(1-q)\,e_2(\omega_2)\,,
\end{equation}
where we used the obvious assumption that the intrinsic probabilities $p_{ij}=e_i(\omega_j)$ are independent of $p$ and $q$.

It is intuitively clear that a machine which randomly selects one of various preparation procedures can be considered as a preparation procedure in itself, thus defining a new state $\omega$. Similarly, a device of randomly selected effects can be interpreted in itself as a new effect $e$. Using these new effects and states the probability~\eref{example} to obtain a 'click' can simply be expressed as $P=e(\omega)$.

It is easy to see that probability~\eref{example} can be recovered by conveniently describing the new states and effects by 
\begin{equation}
\label{eom}
\omega := p \, \omega_1 + (1-p)\, \omega_2 \,,\qquad
e := q \, e_1 + (1-q) \, e_2,
\end{equation}
when we define them to act linearly on each other. We call these new objects \emph{mixed states} and \emph{mixed effects} respectively. The mathematical operations in these terms represent direct physical interpretations. Namely, scalar multiplication refers to being realized only with some probability, whereas addition means the coarse-graining of mutually excluding settings. 

Note that we have introduced probabilistic mixing of devices as a consequence of noise. Of course, this is not the only way to generate mixtures. For example, it could be the experimenter himself who chooses randomly or intentionally between different preparation and measurement devices. 
Thus, with any $p, q$ in the continuous range $[0,1]$ available to an experimenter, probabilistic mixing yields a continuous variety of states and effects that can be realized.

\subsection{Linear spaces, convex combinations, and extremal states and effects}
\label{sec:linear}
%
The previous example shows that it is useful to represent probabilistically mixed states and effects as linear combinations, using the usual rules for addition and multiplication of probabilities. By doing so, we have represented states and effects as vectors in suitable vector spaces, whose structure, dimension and the choice of the basis we will discuss further below. For now, let us only note that each state $\omega_i$ is represented by a vector in a linear space $V$ and that it is possible to define linear combinations in such a way that vectors coincide if and only if they refer to the same state. Similarly each effect $e_i$ can be represented by a vector in another linear space $V^*$, which is called the \textit{dual space} of $V$. 

The embedding of states and effects in linear spaces allows us to consider arbitrary linear combinations
\begin{equation}
\label{eq:linear}
e=\sum_i\lambda_i\,e_i\,,\qquad\omega=\sum_j\mu_j\,\omega_j
\end{equation}
with certain coefficients $\lambda_i$ and $\mu_j$. Moreover, the fundamental probability table $p_{ij}=e_i(\omega_j)$ induces a bilinear map $V^* \times V \to \mathbb{R}$ by
\begin{equation}
\label{eq:linear2}
e(\omega) = \Bigl[\sum_{i=1}^M\lambda_i\,e_i\Bigr]\Bigl(\sum_{j=1}^N\mu_j\omega_j\Bigr) = \sum_{i=1}^M\sum_{j=1}^N\,\lambda_i\mu_j \,\underbrace{e_i(\omega_j)}_{=p_{ij}}\,,
\end{equation}
generalizing Eq.~(\ref{example}) in the previous example. Note that this bilinear map on $V^*\times V$ should not be confused with an inner scalar product on either $V \times V$ or $V^* \times V^*$. In particular, it does not induce the notion of length, norm, and angles.

At this point it is not yet clear which of the linear combinations in (\ref{eq:linear}) represent physically meaningful objects. However, as shown above, the set of physically meaningful objects will at least include all probabilistic mixtures of the existing states and effects, which are mathematically expressed as \textit{convex combinations} with non-negative coefficients adding up to~$1$. 

States which can be written as convex combinations of other states are referred to as \textit{mixed states}. Conversely, states which cannot be expressed as convex combinations of other states are called \textit{extremal states}. As any convex set is fully characterized by its extremal points, we can reduce the probability table even further by listing only the extremal states, tacitly assuming that all convex combinations are included as well. The same applies to effects.

\subsection{Linear dependencies among extremal states and effects}
%
What is the dimension of the spaces $V$ and $V^*$ and how can we choose a suitable basis? To address these questions it is important to note that the extremal vectors of the convex set of states (or effects) are not necessarily linearly independent. As we shall see below, linear independence is in fact a rare exception that emerges only in classical theories, while any non-classicality will be encoded in certain linear dependencies among the extremal states and effects. 
\begin{table}
\centering
\begin{tabular}{|c||c|c|c|c|c|}
\hline 
{} & $e_1$ & $e_2$ & $e_3$ & $e_4$ & $e_5$ \\ 
\hline 
\hline
$\omega_1$ & 1 & 0 & 1 & 1 & 1  \\ 
\hline 
$\omega_2$ & $\frac{1}{2}$ & 0 & 1 & $\frac{2}{3}$ & $\frac{3}{4}$ \\ 
\hline 
$\omega_3$ & $\frac{1}{2}$ & $\frac{1}{2}$ & 1 & $\frac13$ & $\frac{3}{4}$ \\ 
\hline 
$\omega_4$ & 0 & $\frac{1}{2}$ &  1 & 0 & $\frac{1}{2}$ \\ 
\hline 
\end{tabular} 
\caption{Example of a probability table after removing identical columns and rows.}
\label{tab:data1}
\end{table}

Let us illustrate the construction of a suitable basis in the example of a fictitious model with probabilities listed in Table \ref{tab:data1}. As states and effects are defined as equivalence classes, multiple rows and columns have already been eliminated. However, there are still linear dependencies among the rows and the columns. For example, the effect~$e_5$ is related to the other ones by
\begin{equation}
e_5\;=\;\frac12\,(e_1+e_3)\,.
\end{equation}
Since the expression on the r.h.s. is a convex combination it is automatically assumed to be part of the toolbox so that we can remove the rightmost column from the probability table, obtaining a reduced table in form of a $4 \times 4$ matrix. The remaining (non-convex) linear dependencies are
\begin{equation}
e_4 \;=\; \frac{2}{3}\, e_1- \frac{2}{3}\, e_2 + \frac{1}{3}\, e_3 \,, \qquad \omega_4 \;=\; -\omega_1 + \omega_2 - \omega_3\,.
\end{equation}
so that the rank of the matrix is 3. Since row and column rank of a matrix coincide, the vector spaces $V$ and $V^*$ always have the same dimension
\begin{equation}
\label{eq:equalDimensions}
n := \dim V=\dim V^* = \mbox{rank}[\{p_{ij}\}].
\end{equation}
In other words, the number of different states needed to identify an effect is always equal to the number of different effects needed to identify a state.

As for any vector space representation, there is some freedom in choosing a suitable basis. As for the effects, we may simply choose the first $n$ linearly independent effects ${e_1,\ldots, e_n}$ as a basis of $V^*$, assigning to them the canonical coordinate representation
\begin{equation}
\label{eq:basiseffects}
 e_1 = (1,0,0), \quad e_2 = (0,1,0), \quad e_3 = (0,0,1)\,.
\end{equation}
Likewise we could proceed with the states, choosing ${\omega_1,\omega_2,\omega_3}$ as a basis of $V$, but then the matrix $p_{ij}$ would be quite complicated whenever we compute $e(\omega)$ according to Eq.~(\ref{eq:linear}). Therefore it is more convenient to use the so-called \textit{conjugate basis} $\{{\hat\omega}_1,{\hat\omega}_2,{\hat\omega}_3\}$ which is chosen in such a way that the extremal states are just represented by the corresponding lines in the probability table. In the example given above this means that the states have the coordinate representation
\begin{equation}
\omega_1 = \left(1,0,1\right), \quad \omega_2 = \left(\frac{1}{2}, 0, 1\right), \quad 
\omega_3 = \left(\frac{1}{2}, \frac{1}{2}, 1\right).
\end{equation}
The basis vectors $\{\hat\omega_i\}$ can be determined by solving the corresponding linear equations. In the present example, one can easily show that these basis vectors are given by the (non-convex) linear combinations
\begin{equation}
{\hat\omega}_1 = 2\omega_1-2\omega_2,\quad
{\hat\omega}_2 = -2\omega_2+2\omega_3,\quad
{\hat\omega}_3 = -\omega_1+2\omega_2\,.
\end{equation}
Using the conjugate basis the bilinear map $e(\omega)$ can be computed simply by adding the products of the corresponding  components like in an ordinary Euclidean scalar product.

Recall that the vector spaces $V$ and $V^*$ are \textit{probabilistic} vector spaces which should not be confused with the Hilbert space of a quantum system. For example, probabilistic mixtures cannot be represented by Hilbert space vectors. We will return to this point when discussing specific examples.

\subsection{Reliability}
%
Realistic experiments are not only noisy but also unreliable in the sense that they sometimes fail to produce a result. For example, a preparation procedure may occasionally fail to create a physical object. Similarly, a detector may sometimes fail to detect an incident particle.  

Preparation procedures which create a physical state with certainty are called \textit{reliable}. The same applies to measurement devices which respond to an incident particle with certainty. 

\begin{figure}[t]
\centering\includegraphics[width=125mm]{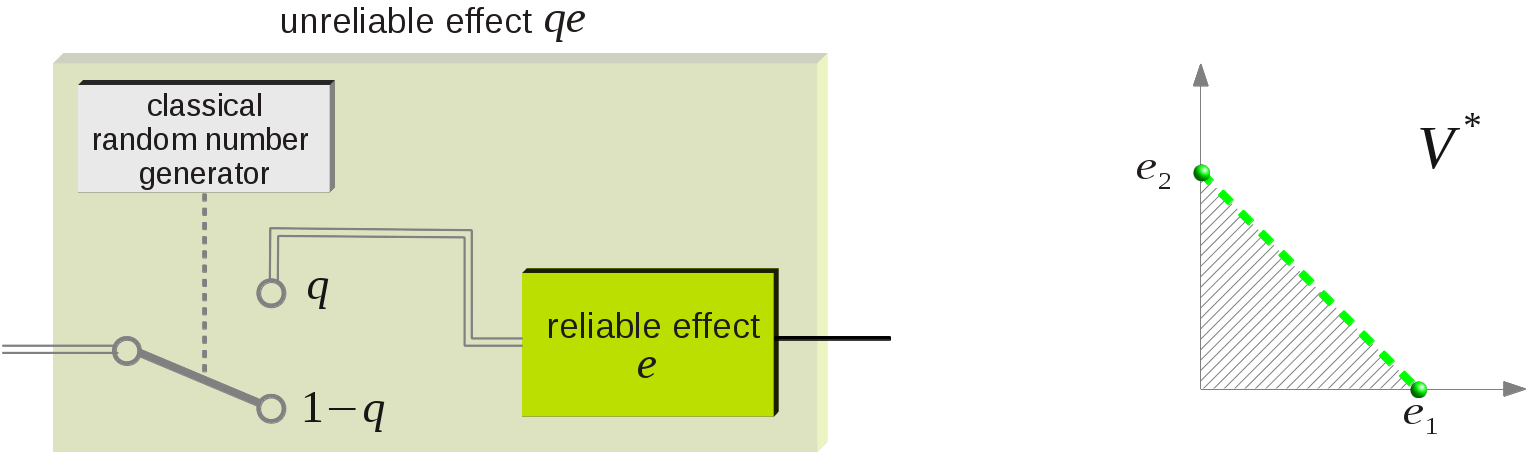}
\caption{Unreliable effects. Left: A reliable effect $e$ can be made unreliable by randomly switching it on and off, constituting a new effect $q\,e$. Right: Reliable effects are represented by points of the convex set in $V^*$ (green dashed line). Including unreliable effects this set is extended to a truncated convex cone (the hatched region) spanned by the extremal effects.}
\label{fig:unreliable}
\end{figure}

An unreliable effect may be thought of as a reliable one that is randomly switched on and off with probability $q$ and $1-q$, as sketched in Fig.~\ref{fig:unreliable}. Applying this effect to a state $\omega$, the probability to obtain a 'click' would be given by $q\,e(\omega)$. This example demonstrates that unreliable effects can consistently be represented as sub-normalized vectors $q\,e \in V^*$ with $0\leq q<1$, extending the set of physical effects to a truncated \textit{convex cone} which is shown as a shaded region in in the right panel of Fig.~\ref{fig:unreliable}. The zero vectors of $V$ and $V^*$ represent the extreme cases of preparation procedures and a measurement apparatuses which always fail to work. 

\subsection{Unit measure and normalization}
%
If a given effect $e$ responds to a specific state $\omega$ with the probability $e(\omega)=1$, then it is of course clear that both the state and the effect are reliable. However, if $e(\omega)<1$, there is no way to decide whether the reduced probability for a 'click' is due to the unreliability of the state, the unreliability of the effect, or caused by the corresponding entry in the probability table. 

To circumvent this problem, it is usually assumed that the toolbox contains a special reliable effect which is able to validate whether a preparation was successful, i.e. it 'clicks' exactly in case of a successful preparation. This effect is called \textit{unit measure} and is denoted by $u$. The unit measure allows us to quantify the reliability of states: If $u(\omega)=1$ the state $\omega$ is reliable, otherwise its rate of failure is given by $1-u(\omega)$. 

The unit measure can be interpreted as a norm 
\begin{equation}
\label{StateNorm}
||\omega|| = u(\omega) 
\end{equation}
defined on states in the convex cone of~$V$. By definition, the normalized states with $u(\omega)=1$ are just the reliable ones. The corresponding set (the green dashed line in Fig.~\ref{fig:unreliable}) is usually referred to as the \textit{state space} $\Omega$ of the theory. 

In the example of Table~\ref{tab:data1} it is easy to see that the effect $e_3$ plays the role of the unit measure. Since the unit measure cannot be represented as a convex combination of other effects, it is by itself an extremal effect and thus may be used as a basis vector of $V^*$. Here we use the convention to sort the Euclidean basis of $V^*$ in such a way that the unit measure appears in the last place, i.e. $e_M\equiv u$. Using this convention the norm of a state is just given by its last component. For example, in Table~\ref{tab:data1}, where $e_3=u$, the third component of all states $\omega_1,\ldots,\omega_4$ is equal to $1$, hence all states listed in the table are normalized and thus represent reliable preparation procedures.

The unit measure also induces a norm on effects defined by
\begin{equation}
\label{EffectNorm}
||e|| = \max_{\omega\in\Omega} e(\omega)\,.
\end{equation}
Since $e(\omega)\leq 1$ an effect is normalized (i.e. $||e||=1$) if and only if there exists a state $\omega$ for which $\omega(e)=1$. By definition, such an effect is always reliable. The opposite is not necessarily true, i.e. reliable effects may be non-normalized with respect to the definition in (\ref{EffectNorm}). 

Note that a `unit state', analogous to the unit effect $u$, is usually not introduced since this would correspond to a preparation procedure to which \textit{every} reliable effect of the toolbox responds with a '1' with certainty, which is highly unphysical. If we had introduced such a `unit state', it would have allowed us to define a norm on effects analogous to Eq.~(\ref{StateNorm}), preserving the symmetry between states and effects. Using instead the norm (\ref{EffectNorm}) breaks the symmetry between the spaces $V$ and $V^*$. 

As we will see in the following, the unit measure $u$ plays a central role in the context of consistency conditions and it is also needed to define measurements with multiple outcomes. Moreover, the definition of subsystems in Sect. \ref{sec:subsystems} relies on the unit measure. 

\subsection{General consistency conditions}
\label{sec:consistency}
%
The concepts introduced so far represent only the factual experimental observations and immediate probabilistic consequences. However, the purpose of a physical model is not only to reproduce the existing data but rather to make new predictions, eventually leading to a set of hypotheses that can be tested experimentally.

In order to give a GPT the capability of making new predictions one has to postulate additional extremal states and effects which are not yet part of the existing toolbox. Such an extension is of course not unique, rather there are various possibilities which can be justified in different ways. For example, a particular extension might be reasonable in view of the underlying structure and the expected symmetries of the physical laws. Moreover, certain expectations regarding the relationship between the parameters of the apparatuses and the corresponding states and effects as well as analogies to other models could inspire one to postulate a specific structure of the state space and the set of effects. This includes dynamical aspects of the systems, which are absorbed into preparations and measurements in the present framework.

\begin{figure}[t]
\begin{center}
\includegraphics[width=50mm]{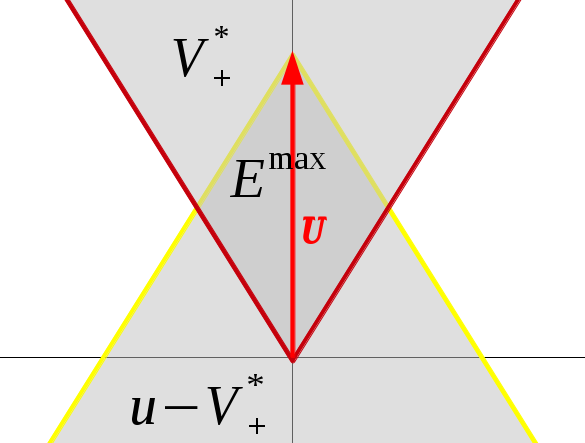} \hspace{10mm}
\includegraphics[width=50mm]{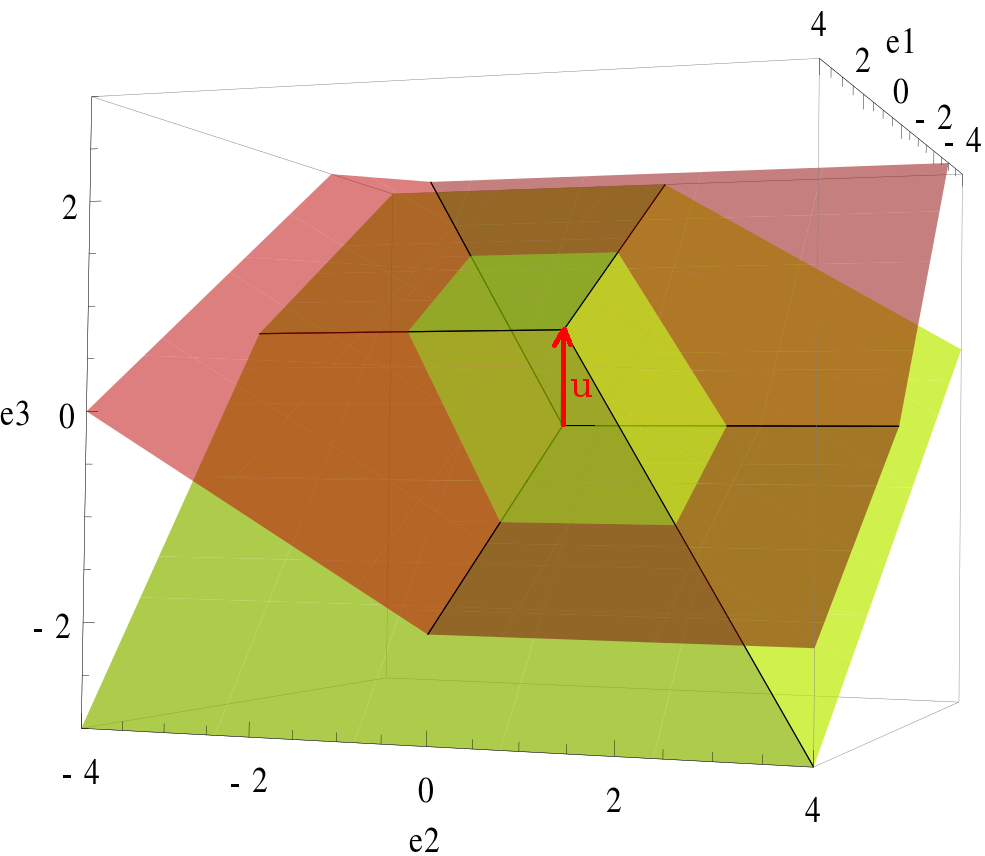}
\end{center}
\caption{Consistency conditions. Left: Schematic illustration of the lower and the upper bound, defining the intersection $E^\mathrm{max}$. Right: The same construction for the probabilities listed in Table~\ref{tab:data1} in the three-dimensional representation~(\ref{eq:basiseffects}). The red (yellow) planes indicate the lower (upper) bound. The maximal set of effects $E^\mathrm{max}$ is the enclosed parallelepiped in the center.}
\label{fig:constraints}
\end{figure}

However, not every extension of states and effects gives a consistent theory. First of all, the extension should be introduced in such a way that any combination of effects and states yields a probability-valued result, i.e., 
\begin{equation}
 \label{eq:probrange}
 0 \leq e(\omega) \leq 1 \quad \forall e \in E, \omega \in \Omega.
\end{equation}
This restriction consists of a lower and an upper bound. The lower bond $0 \leq e(\omega)$, the so-called \textit{non-negativity constraint}, remains invariant if we rescale the effect $e$ by a positive number. In other words, for any effect $e$ satisfying the non-negativity constraint, the whole positive ray $\lambda \, e$ with $\lambda \geq 0$ will satisfy this constraint as well. The set of all rays spanned by the non-negative effects is the so-called \emph{dual cone}, denoted as
\begin{equation}
\label{eq:dualcone}
V^*_+ := \left\{ e \in V^*\,|\, e(\omega) \geq 0 \, \forall \omega \in \Omega \right\}.
\end{equation}
The upper bound can be expressed conveniently with the help of the unit measure $u$. Since the unit measure is the unique effect giving $1$ on all normalized states, it is clear that  $e(\omega)\leq 1$ if and only if $u(\omega)-e(\omega)=(u-e)(\omega)\geq 0$, i.e., the complementary effect $u-e$ must be included in the dual cone given by \eref{eq:dualcone}. Note that this criterion is valid not only for normalized states but also for sub-normalized states. This means that the set of effects, which obey the upper bound $e(\omega)\leq 1$, is just $u-V_+^*$. Consequently, the set which satisfies both bounds in \eref{eq:probrange}, is just the intersection of $V^*_+$ and $u-V^*_+$, as illustrated in Fig.~\ref{fig:constraints}. This maximal set of effects is denoted by\footnote{In the literature this set is also denoted by $ [\emptyset, u]$ because of a partial ordering induced by $V^*_+$, as we explain in more detail in appendix \ref{app:partialorder}.}
\begin{equation}
\label{eq:maxset}
E^\mathrm{max} \;=\; V^*_+ \cap \left(u - V^*_+\right) .
\end{equation}
Thus, if we extend the theory by including additional effects, the resulting set of effects $E$ has to be a subset of this maximal set, i.e.
\begin{equation}
E \subseteq E^\mathrm{max} .
\end{equation}
A theory that includes the full set $E^\mathrm{max}$ of effects is referred to as satisfying the \textit{no-restriction hypothesis} \cite{norestriction}. It can be shown that classical probability theory and quantum theory both satisfy the no-restriction hypothesis, but in general there is no reason why the preparations in our current toolbox should fully determine the range of possible measurements. Note that for consistency the special effects $\emptyset$ and the unit measure $u$ have to be included. In addition, for any effect $e\in E$ the complement $\bar{e} = u - e$ needs to be included as well.

Similarly we may extend the theory by including additional states. Here we have to specify the set of states which satisfy \eref{eq:probrange} for a given set of effects $E$. Generally the inclusion of additional states imposes additional restrictions on possible effects and vice versa. Consequently, there is a trade-off between states and effects whenever a theory is extended without changing the dimension of the vector spaces $V$ and $V^*$.

A given GPT can also be generalized by increasing the dimension of $V$ and $V^*$. In fact, as will be shown in Sect. \ref{sec:nonclassicalbyrestriction}, every non-composite system from an arbitrary GPT can equivalently be realized as a classical theory in a higher-dimensional state space combined with suitable restrictions on the effects. However, as we will see in Sect. \ref{sec:jointsystems}, the treatment of multipartite systems leads to additional consistency conditions which cannot be fulfilled by restricted classical systems in higher dimensions, allowing us to distinguish classical from genuine non-classical theories.

\subsection{Jointly measurable effects}
\label{sec:jointmeas}
%
A set of effects is said to be \textit{jointly measurable} if all of them can be evaluated in a single measurement, meaning that there exists a measurement apparatus that contains all these effects as marginals. By definition, effects belonging to the same measurement apparatus are jointly measurable. However, a GPT may also include effects that cannot be measured jointly. Therefore, it is of interest to formulate a general criterion for joint measurability.

Before doing so, let us point out that joint measurability neither requires the effects to be evaluated at the same time nor does it mean that they do not influence each other. For example, let us consider a non-destructive measurement of effects $\{e^1_i\}$ with results $\{\chi^1_j\}$ followed by a second measurement. The results $\{\chi^2_j\}$ of the second measurement correspond to effects $e^2_j$ with the proviso that the first measurement has already  been carried out. If the first measurement was not carried out, we would obtain potentially different effects. Nevertheless, the whole setup measures all effects $\{e^1_i\}$ and $\{e^2_j\}$ jointly, irrespective of the fact that the second group depends on the first one.  

Joint measurability of effects is in fact a weaker requirement than non-disturbance and commutativity of measurements. In standard quantum theory these terms are often  erroneously assumed to be synonyms. This is because in the special case of projective measurements they happen to coincide. However, as shown in \cite{heinosaari10,reeb13}, they even differ in ordinary quantum theory in the case of non-projective measurements.

Let us now formally define what joint measurability means. Consider two effects $e_i$ and $e_j$. Applied to a state $\omega$ each of them produces a classical one-bit result $\chi_i\in\{0,1\}$ and $\chi_j\in\{0,1\}$. Joint measurability means that there exists another single measurement apparatus in the toolbox that allows us to extract two bits $(\tilde\chi_i,\tilde\chi_j)$ by Boolean functions with the same measurement statistics as $(\chi_i,\chi_j)$.

In other words, two effects $e_i$, $e_j$ are jointly measurable if the toolbox already contains all effects which are necessary to set up the corresponding Boolean algebra, i.e. there are mutually excluding effects $e_{i \wedge j}, e_{i \wedge \overline{j}}, e_{\overline{i} \wedge j}, e_{\overline{i} \wedge \overline{j}}$ with the properties
\begin{eqnarray}
\nonumber
e_i &= e_{i \wedge j} + e_{i \wedge \overline{j}}\,, \qquad e_j = e_{i \wedge j} + e_{\overline{i} \wedge j}\\
\label{eq:logicalops}
u &= e_{i \wedge j} + e_{i \wedge \overline{j}} + e_{\overline{i} \wedge j} + e_{\overline{i} \wedge \overline{j}}\\
\nonumber
e_{i \vee j} &= e_i + e_j - e_{i \wedge j}\,.
\end{eqnarray}
Let us use Eqs. \eref{eq:logicalops} to rewrite $e_{i \wedge j}$ in three different ways:
\begin{eqnarray}
\nonumber
 e_{i \wedge j} &= e_i - e_{i \wedge \overline{j}}\\
\label{eq:ANDvariations}
  &= e_j - e_{\overline{i} \wedge j}\\
\nonumber
&= e_i + e_j - u + e_{\overline{i} \wedge \overline{j}} \,\,.
\end{eqnarray}
We can now translate the joint measurability condition to
\begin{equation}
 \exists e_1, e_2, e_3, e_3 \in E: \,\,\, e_1 \,=\, e_i - e_2 \,=\, e_j - e_3 \,=\, e_i + e_j - u + e_4\,.
\end{equation}
This condition can be rewritten elegantly as an intersection of sets
\begin{equation}
 \label{eq:ANDset}
 E \cap (e_i - E) \cap (e_j - E) \cap (e_i + e_j - u + E) \neq \{\}.  
\end{equation}
For joint measurability of the effects $e_i$, $e_j$ this set has to be non-empty. If this is the case, any choice of the AND effect $e_{i \wedge j}$ in the intersection (\ref{eq:ANDset}) allows one to consistently construct all other effects by means of Eqs.~\eref{eq:logicalops}. This means that joint measurability of two effects can be implemented in various ways with different $e_{i \wedge j}$. Note that the status of joint measurability of a given set of effects may even change when a theory is extended by including additional effects.

\subsection{Complete and incomplete Measurements}
%
A \textit{measurement} is defined as a set of jointly measurable effects. If these effects have a non-trivial overlap $e_{i \wedge j} \neq 0$ we can further refine the measurement by including the corresponding AND effects. Thus, we can describe any measurement by a set of mutually excluding effects $\{e_k\}$, where only one of the outcomes $\chi_k$ occurs, as sketched in Fig.~\ref{fig:expsetup}. These refined effects have no further overlap, i.e. $e_{k \wedge l} = 0$ for $k \neq l$. Moreover, these effects can be coarse-grained  by computing their sum $e_{k \vee l} = e_k + e_l$. 

A measurement is called \textit{complete} if all mutually excluding effects sum up to the unit measure $u$. Obviously an incomplete measurement can be completed by including a failure effect $e_m = u - \sum_{i=1}^{m-1} e_i$ that is complementary to all other effects. As a consequence a complete measurement maps a normalized state to a normalized probability distribution.

\section{Examples}

\subsection{Classical probability theory}
%
Classical systems have properties that take definite perfectly distinguishable values that can be directly revealed via measurements. Probabilistic mixtures can be regarded as a mere consequence of subjective ignorance.

\begin{figure}[t]
\centering\includegraphics[width=90mm]{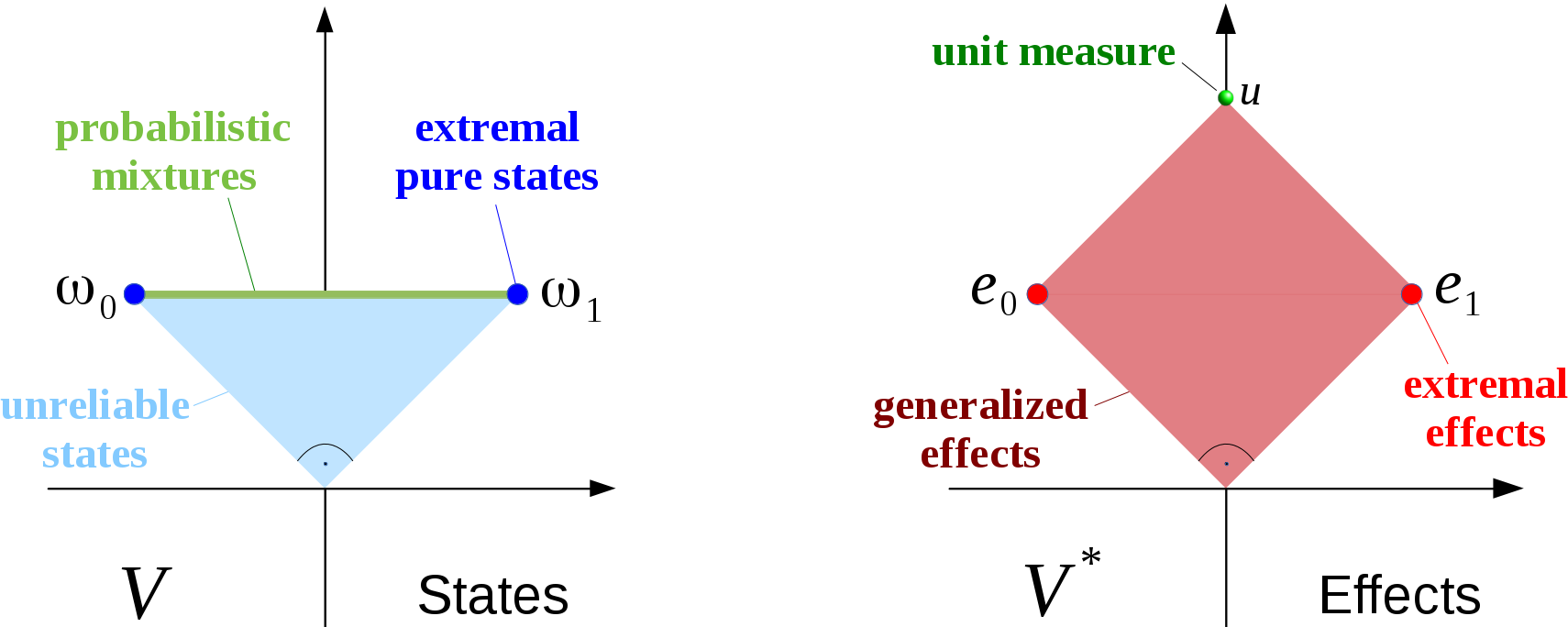}
\caption{State and effect space of a classical bit  in the GPT formalism with the probability table $e_i(\omega_j)=\delta_{ij}$. In classical systems the extremal states and effects are linearly independent and can be used as an orthonormal basis of the vector spaces.}
\label{fig:bit}
\end{figure}

In the GPT framework the different possible definite values of a classical system are represented by the pure states $\omega_i$. They are linearly independent and can be used as an Euclidean basis of the linear space $V$. The corresponding state space is a probability simplex (see Fig.~\ref{fig:bit}). Probabilistic mixtures are represented by convex combinations of pure states. As the pure states form a basis, any mixed state can be uniquely decomposed into pure states weighted by the probabilities of occurrence.

The perfect distinguishability of pure states means that the extremal effects $e_j$ simply read out whether a particular value has been realized or not, i.e. $e_j(\omega_i) = \delta_{ij}$. Like the pure states in $V$ these effects provide an Euclidean basis for $V^*$. Furthermore, the zero effect $\emptyset$, and coarse-grained basis effects $e_j$ have to be included as additional extremal effects. In particular, this includes the unit measure $u$ which is obtained by coarse-graining all basis effects $e_j$. The unit measure responds with a '$1$' to any successful preparation of a classical system, independent of its values. In classical systems all effects are jointly measurable.\\

\subsection{Standard quantum theory: State space}
%
Most textbooks on quantum theory introduce quantum states as vectors $\ket{\Psi}$ of a complex Hilbert space $\mathcal{H}$. These vectors represent pure quantum states. The existence of a Hilbert space representation is in fact a special feature of quantum mechanics. In particular, it allows one to combine any set of pure states $\ket{\Psi_i}$ linearly by coherent superpositions
\begin{equation}
\label{QuantumSuperposition}
\ket{\phi}=\sum_i\lambda_i \ket{\psi_i}\,, \qquad \lambda_i \in \mathbb C\,, \qquad \sum_i |\lambda_i|^2=1\,.
\end{equation}
Note that the resulting state $\ket\phi$ is again a pure state, i.e. coherent superpositions are fundamentally different from probabilistic mixtures. In fact, Hilbert space vectors alone cannot account for probabilistic mixtures. 

To describe mixed quantum states one has to resort to the density operator formalism. To this end the pure states $\ket{\Psi}$ are replaced by the corresponding projectors $\rho_\Psi = \ket{\Psi}\bra{\Psi}$. Using this formulation one can express probabilistically mixed states as convex combinations of such projectors, i.e.  
\begin{equation}
\label{QuantumDensityMatrix}
\rho = \sum_i p_i \ket{\Psi_i}\bra{\Psi_i}\,,\qquad \sum_i p_i=1\,.
\end{equation}
As the expectation value of any observable $\mathbf A$ is given by $\tr[\rho \mathbf A]$, it is clear that the density matrix includes all the available information about the quantum state that can be obtained by means of repeated measurements. 

It is important to note that the density matrix itself does not uniquely determine the $p_i$ and $\ket\psi_i$ in (\ref{QuantumDensityMatrix}), rather there are many different statistical ensembles which are represented by the same density matrix. For example, a mixture of the pure qubit states $\ket{0}\bra{0}$ and $\ket{1}\bra{1}$ with equal probability, and a mixture $\ket{+}\bra{+}$ and $\ket{-}\bra{-}$ of the coherent superpositions $\ket{\pm} = \frac{1}{\sqrt{2}} \, (\ket{0} \pm \ket{1})$ are represented by the same density matrix
\begin{equation}
\rho = \frac12 \, \Bigl(\ket{0}\bra{0} + \ket{1}\bra{1}\Bigr) = \frac12 \, \Bigl(\ket{+}\bra{+} + \ket{-}\bra{-}\Bigr)\,,
\end{equation}
meaning that these two ensembles cannot be distinguished experimentally. Thus, in ordinary quantum mechanics the density matrices $\rho$ label equivalence classes of indistinguishable ensembles and therefore correspond to the physical states $\omega$ in the GPT language. The set of all quantum states (including probabilistic mixtures) can be represented by Hermitean matrices with semi-definite positive eigenvalues. A state is normalized if $\tr[\rho] = 1$, reproducing the usual normalization condition $\braket{\psi}{\psi} = 1$ for pure states.

Identifying the density operators as states, one faces the problem that these operators live in a complex-valued Hilbert space whereas the GPT framework introduced above involves only real-valued vector spaces. In order to embed quantum theory in the GPT formalism, let us recall that a $n \times n$ density matrix can be parametrized in terms of $SU(n)$ generators with real coefficients. For example, the normalized density matrix of a qubit can be expressed in terms of $SU(2)$-generators (Pauli matrices) as
\begin{equation}
\rho= \frac12 \, (\mathbf{1} + a \, \sigma^x + b \, \sigma^y + c \, \sigma^z)
\end{equation}
with real coefficients $a,b,c \in [-1,1]$ obeying the inequality $a^2+b^2+c^2 \leq 1$.  Regarding the coefficients $(a,b,c)$ as vectors in $\mathbb{R}^3$, the normalized states of a qubit form a ball in three dimensions. The extremal pure states are located on the surface of this ball, the so-called \textit{Bloch sphere}. 

\begin{figure}[t]
\centering\includegraphics[width=100mm]{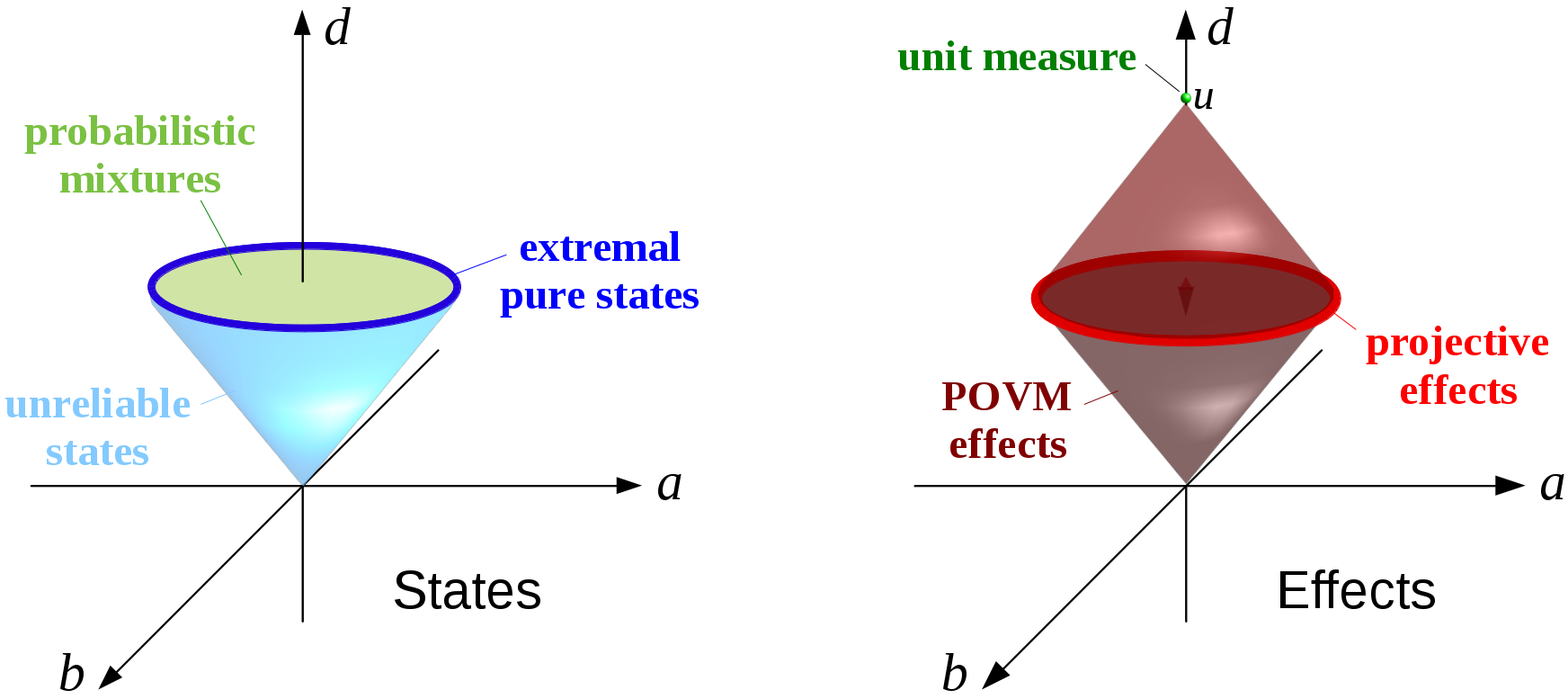}
\caption{State and effect spaces of a quantum-mechanical qubit in the GPT formalism. Since the vector space are four-dimensional the figure shows a three-dimensional projection, omitting the third coefficient $c$. }
\label{fig:qubit}
\end{figure}

In order to include non-normalized states (e.g. unreliable preparation procedures), we have to append a forth coefficient $d$ in front of the unit matrix, i.e.
\begin{equation}
\rho= \frac12 \, (d \,\mathbf{1} + a \, \sigma^x + b \, \sigma^y + c \, \sigma^z)
\end{equation}
which is $1$ for any normalized state and less than $1$ if the preparation procedure is unreliable. The four coefficients $(a,b,c,d)$ provide a full representation of the state space in $ {\mathbb R}^4$ according to the GPT conventions introduced above. This state space is illustrated for the simplest case of a qubit in the left panel of Fig.~\ref{fig:qubit}.

\subsection{Standard quantum theory: Effect space}
%
As there are pure and mixed quantum states there are also two types of measurements. Most physics textbooks on quantum theory are restricted to `pure' measurements, known as projective measurements. A projective measurement is represented by a Hermitean operator $\mathbf A$ with the spectral decomposition
\begin{equation}
\mathbf A = \sum_a a\,\ket a \bra a
\end{equation}
with real eigenvalues $a$ and a set of orthonormal eigenvectors $\ket a$. If such a measurement is applied to a system in a pure state $\ket\psi$ it collapses onto the state $\ket a$ with probability $p_a=|\braket{a}{\psi}|^2$. Introducing projection operators $E_a = \ket a \bra a$ and representing the pure state by the density matrix $\rho=\ket \psi \bra \psi$ this probability can also be expressed as
\begin{equation}
p_a = |\braket{a}{\psi}|^2 = \braket{a}{\psi}\braket{\psi}{a} = \tr[E_a^\dagger \, \rho],
\end{equation}
i.e. the absolute square of the inner product between bra-ket vectors is equivalent to the Hilbert-Schmidt inner product of operators $E_a$ and $\rho$. Hence we can identify the projectors $E_a = \ket a \bra a$ with extremal effects in the GPT framework, where $e_a(\omega) = \tr[E_a\rho]$. As the projectors $E_a$ cannot be written as probabilistic combinations of other projectors, it is clear that they represent extremal effects. As all these effects sum up to $\sum_a E_a=1$, the unit measure $u$ is represented by the identity matrix. 

Turning to generalized measurements, we may now extend the toolbox by including additional effects which are defined as probabilistic mixtures of projection operators of the form 
\begin{equation}
\label{GeneralizedEffects}
E_a = \sum_i q_i \ket{a_i}\bra{a_i} \,, \qquad 0\leq q_i \leq 1.
\end{equation}
As outlined above, such mixtures can be thought of as unreliable measurements. A general measurement, a so-called \textit{positive operator valued measurement} (POVM), consists of a set of such effects that sum up to the identity. 

Interestingly, the generalized effects in Eq.~(\ref{GeneralizedEffects}) are again positive operators. i.e. mixed effects and mixed quantum states are represented by the same type of mathematical object. Therefore, quantum theory has the remarkable property that the spaces of states and effects are isomorphic. In the GPT literature this special property is known as (strong) \textit{self-duality}.

Note that for every given pure state $\rho = \ket{\Psi}\bra{\Psi}$ there is a corresponding measurement operator $E = \ket{\Psi}\bra{\Psi}$ that produces the result $\tr[E\,\rho]=1$ with certainty on this state. In so far the situation is similar as in classical systems. However, in contrast to classical systems, it is also possible to obtain the same outcome on other pure states with some probability. This means that in quantum mechanics pure states are in general not perfectly distinguishable.

\subsection{The gbit}
%
\begin{figure}[t]
\centering\includegraphics[width=100mm]{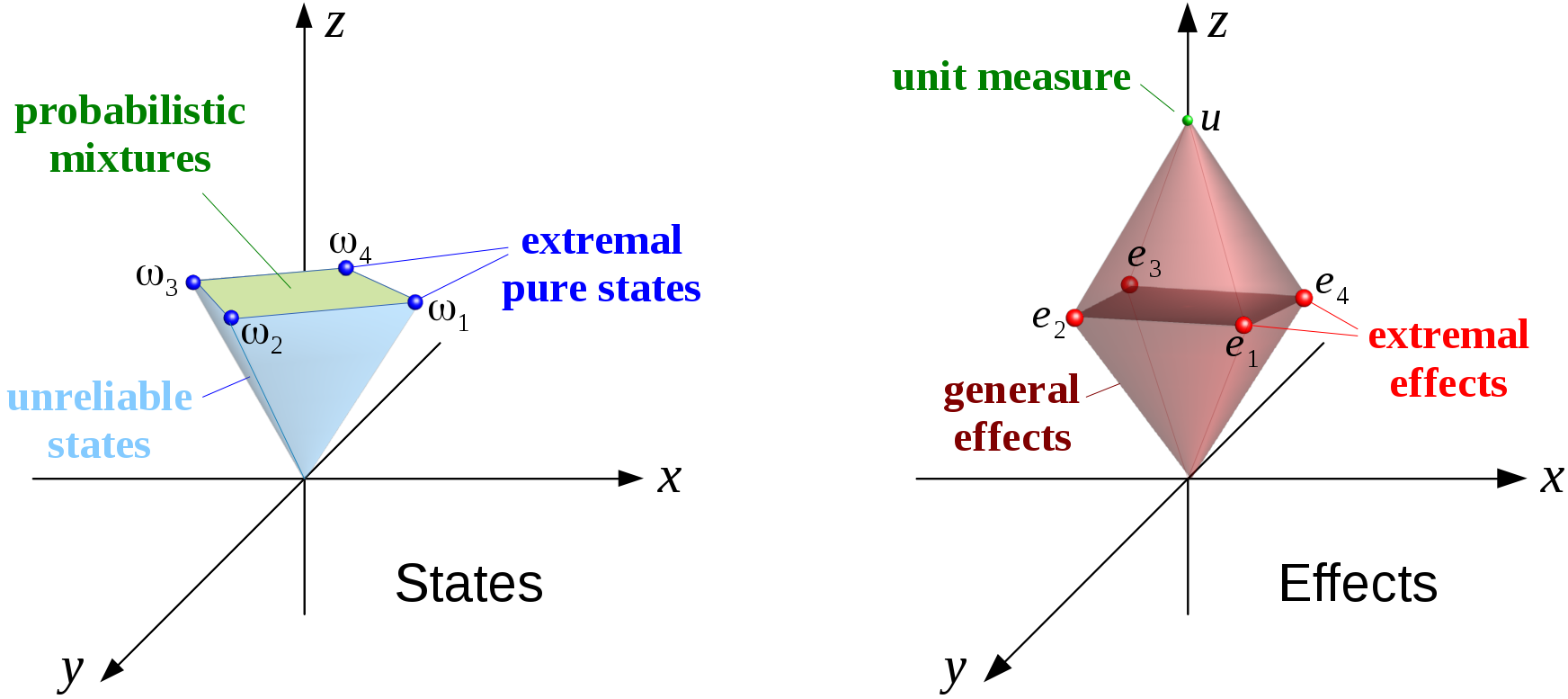}
\caption{State and effect spaces of a gbit. }
\label{fig:gbit}
\end{figure}
A popular toy theory in the GPT community, which is neither classical nor quantum, is the \textit{generalized bit}, the so-called gbit. This theory has a square-shaped state space defined by the convex hull of the following extremal states $\omega_i$:
\begin{equation}
\omega_1 = (1,0,1) ,\quad  \omega_2 = (0,1,1),\quad \omega_3 = (-1,0,1), \quad  \omega_4 = (0,-1,1)\,.
\end{equation}
The corresponding set of effects usually includes all linear functionals that give probability-valued results when applied to states. This means that the effect space is given by the convex hull of the following extremal vectors:
\begin{eqnarray}
 e_1 = \frac12 \,(1,1,1) \quad & e_2 = \frac12 \,(-1,1,1),\nonumber \\
 e_3 = \frac12 \,(-1,-1,1), \quad & e_4 = \frac12 \,(1,-1,1)
\end{eqnarray}
together with the zero effect and the unit measure
\begin{equation}
 \emptyset = (0,0,0), \quad  u = (0,0,1).
\end{equation}
Remarkably, in contrast to the classical and the quantum case, each extremal effect $e_i$ gives certain outcomes on more than one extremal state.

\subsection{Other toy theories}

A whole class of toy theories with two-dimensional state spaces given by regular polygons was introduced in \cite{locallimits}. Remarkably, these include some of the above standard cases, which correspond to state spaces with a particular number $n$ of vertices. The state space of a classical theory with three distinguishable pure states is given by a triangle-shaped state space, that is the regular polygon with three vertices ($n=3$). The square shaped state space of the gbit corresponds to $n=4$. In the limit $n \to \infty$, however, we get a two-dimensional subspace of a qubit, inheriting some of the quantum features. The polygon theories can therefore be used to compare the different standard cases. Furthermore, increasing the number of vertices yields a transition from a classical theory and a gbit to a quantum-like theory in the limit of infinitely many vertices.
A continuous transition between a classical system and a gbit was studied by a different class of toy theories \cite{generalbox}. It consists of a two-dimensional state space with four vertices. The location of one of the vertices is parametrized, such that the square and the triangle appear as special cases for particular parameters. 

Further examples of state spaces discussed in the literature include a complicated three-dimensional cushion-like state space to model three-slit interference \cite{ududecthesis}, a cylinder-shaped state space \cite{typicalentanglement}, a house-shaped state space \cite{locallimits}, hypersheres of arbitrary dimensions used as a generalization of the Bloch sphere of qubits \cite{hyperbit}, a three-dimensional state space with triangle-shaped and disc-shaped subspaces \cite{barnum10-2} and a three-dimensional approximation of the Bloch ball with finite extremal states \cite{pfisternature}.

All the theories introduced so far, include the full set of potential effects. That is, they obey the no-restriction hypothesis. Toy theories with restricted effect sets are discussed in \cite{norestriction}. Particular interesting examples in this work are theories with inherent noise and a construction that mimics the state-effect duality of quantum theory by restricting the effect set of general theories. Another popular example of a restricted GPT is the probabilistic version of Spekken's toy theory \cite{spekkens07}, given by octrahedron-shaped state space and effect set \cite{garner13, norestriction}.

\subsection{Special features of quantum theory}
\label{sec:QTsinglefeatures}

Having introduced some examples of GPTs let us now return to the question what distinguishes quantum mechanics as the fundamental theory realized in Nature from other possible GPTs. Although a fully conclusive answer is not yet known, one can at least identify various features that characterize quantum mechanics as a particularly elegant theory.

\subsubsection{Continuous state and effect spaces:}
Comparing the state and effect spaces in Figs. \ref{fig:qubit} and \ref{fig:gbit} visually, one immediately recognizes that quantum theory is special in so far as extremal states and effects form \textit{continuous manifolds} instead of isolated points. In the Hilbert space formulation this 
allows one to construct coherent superpositions and to perform reversible unitary transformations, giving the theory a high degree of symmetry. Note that coherent superpositions and probabilistic mixtures are very different in character: While mixtures exist in all GPTs, coherent superpositions turn out to be a special property of quantum theory. GPTs do in general not admit reversible transformations between different extremal states which would be a prerequisite for the possibility of superpositions~\cite{spekkens07, garner13}. 

Quantum theory does not only allow one to relate pure states by reversible unitary transformations (transitivity) \cite{hardy01, dakic11, masanes11}, but even mixed states can be reversibly transformed into each other (homogeneity) \cite{wilce12}. Moreover, the continuous manifold of infinitely extremal quantum states  does not require infinite-dimensional vector spaces. For example, the state space of a qubit (Bloch ball) is three-dimensional although it has infinitely many extremal points. 

\subsubsection{Distinguishability and sharpness:}
The possibility of reversible transformations between extremal states has direct consequences in terms of the information processing capabilities of a theory \cite{barrett07}. As we have seen, in non-classical theories pairs of states are in general not perfectly distinguishable. Remarkably, quantum theory is also special in so far as it allows for a weaker notion of perfect distinguishability \cite{chiribella11}, namely, for any extremal state one can find a certain number of other perfectly distinguishable extremal states (the orthogonal ones in the Hilbert space formulation). This number is called the \textit{information capacity} of the system which corresponds to the classical information that can be encoded in such a subsystem of distinguishable states. In quantum theory it is equal to the dimension of the Hilbert space. 

Obviously, any GPT with given state and effect spaces has well-defined subsets of perfectly distinguishable states and therewith a well-defined information capacity. Remarkably, for quantum theory the opposite is also true, i.e., the information capacity of a system can be shown to determine its state space \cite{hardy01, dakic11}. As a consequence it turns out that a system of given information capacity includes non-classical ones with a lower information capacity as subspaces \cite{masanes11}, which allows for an ideal compression of the encoded information \cite{chiribella11}. This embedding is reflected by a rich geometrical structure of state spaces that is still to a large extent unexplored. The interested reader may be referred to~\cite{qgeom} for an detailed discussion of the geometry of quantum state spaces. An example of a state space that is still comparably low dimensional, but nevertheless has a highly non-trivial structure, is the \textit{qutrit}, a quantum system with information capacity three. It has extremal points that lie on the surface of an eight-dimensional ball with radius $\sqrt{2/3}$. However, the sphere is only partially covered with extremal states. In particular, for any pure state of a qutrit there is a subspace with information capacity 2 including all states that can be perfectly distinguished from the first one. As quantum systems with information capacity 2 are represented by the three-dimensional Bloch ball, we can conclude that there is an opposing Bloch-ball-shaped facet for any extremal point of the qutrit state space. 

Quantum theory is also special in so far as for any extremal state $\omega$ there exists a unique extremal effect $e$ which gives $e(\omega)=1$ while it renders a strictly lower probability for all other extremal states. Therefore, this effect allows one to unambiguously identify the state $\omega$. The existence of such identifying effects is another special quantum feature known as sharpness \cite{hardy11}. 

\subsubsection{Strong self-duality:}
Another striking feature of quantum theory is the circumstance that extremal states and the corresponding identifying effects are represented by the same density operator. This is related to the fact that quantum theory is (strongly) self-dual, i.e. the cone of non-normalized states and its dual cone coincide \cite{teleportation} and obey the no-restriction hypothesis \cite{norestriction}. It was shown that this is a consequence of bit symmetry, i.e. all pairs of distinguishable states can be reversibly transformed into each other \cite{selfduality}.\\

To summarize, from the perspective of GPTs quantum theory has remarkable characteristic properties which may give us an idea why this theory is the one realized in Nature. On the other hand, various other features that seem special for quantum theory turned out to be common for non-classical theories within the GPT framework. Examples are the operational equivalence of different ensembles and the impossibility to clone a state \cite{masanes06}.

Note, that so far we have only discussed single (i.e. non-composite) systems. Already on this level quantum theory exhibits very special features that are hard to find in any other toy theories. Nevertheless, as we will show in the next section, any non-classical single theory can be simulated by a higher dimensional classical system with appropriate restrictions. 

\section{Non-classicality by restriction}

\label{sec:nonclassicalbyrestriction}
Any non-classical single (non-partitioned) system can be interpreted as a classical system with appropriately restricted effects in higher dimensions. This can easily be illustrated in the example of Table~\ref{tab:data1}. Suppose we extend the table by one additional column for each preparation procedure $\omega_i$ which contains a '1' for $\omega_i$ and '0' otherwise (see Table~\ref{tab:extended}). Obviously, these additional columns can be interpreted as additional effects that allow us to perfectly distinguish different preparation procedures, just in the same way as in a classical model. In other words, by adding these columns we have extended the model to a classical one in a higher-dimensional space, where each of the preparation procedures is represented by a different pure state. The original effects can simply be interpreted as coarse-grained mixtures of the additional effects. 

\begin{table}
\centering
\begin{tabular}{|c||c|c|c|c|c||c|c|c|c|}
\hline 
{} & $e_1$ & $e_2$ & $e_3$ & $e_4$ & $e_5$ & $e_6$ & $e_7$ & $e_8$ & $e_9$ \\ 
\hline 
\hline
$\omega_1$ & 1 & 0 & 1 & 1 & 1 & 1 & 0 & 0 & 0\\ 
\hline 
$\omega_2$ & $\frac{1}{2}$ & 0 & 1 & $\frac{2}{3}$ & $\frac{3}{4}$  & 0 & 1 & 0 & 0\\ 
\hline 
$\omega_3$ & $\frac{1}{2}$ & $\frac{1}{2}$ & 1 & $\frac13$ & $\frac{3}{4}$  & 0 & 0 & 1 & 0\\ 
\hline 
$\omega_4$ & 0 & $\frac{1}{2}$ &  1 & 0 & $\frac{1}{2}$  & 0 & 0 & 0 & 1\\ 
\hline 
\end{tabular} 
\caption{Copy of Table \ref{tab:data1} extended by four additional effects $e_6,e_7,e_8,e_9$, converting the non-classical theory into a classical one in higher dimensional space.}
\label{tab:extended}
\end{table}

Conversely, it is also possible to restrict a classical system in such a way that it seems to acquire non-classical features. Such an example was given by Holevo in 1982 \cite{holevo82}: Take a classical system with four pure states 
\begin{equation}
\fl \qquad
\omega_1 = (1,0,0,0), \quad  \omega_2 = (0,1,0,0),\quad
\omega_3 = (0,0,1,0), \quad  \omega_4 = (0,0,0,1)
\end{equation}
representing four distinguishable values. These extremal states span a three-dimensional tetrahedron of normalized mixed states embedded in four-dimensional space. The corresponding extremal effects are given by the vertices of the four-dimensional hypercube $e=(x_1,x_2,x_3,x_4)$ with $x_i \in \{0,1\}$, including the zero effect $\emptyset = (0,0,0,0)$ and the unit measure $u = (1,1,1,1)$. By definition, two states $\omega = (y_1, y_2, y_3, y_4)$ and $\omega' = (y'_1, y'_2, y'_3, y'_4)$ are operationally equivalent whenever 
\begin{equation}
e(\omega)=e(\omega') \quad \Leftrightarrow \quad
\sum_{i=1}^4 x_i \, y_i = \sum_{i=1}^4 x_i \, y'_i 
\end{equation}
for all available effects $e$, which in this case means that all components $y_i=y_i'$ coincide.

Now, let us restrict our toolbox of effects to a subset where
\begin{equation}
\label{restriction}
x_1 + x_2 = x_3 + x_4\,.
\end{equation}
As a result, $\omega$ and $\omega'$ can be operationally equivalent even if the components $y_i$ and $y_i'$ are different. More specifically, if there is a $t\neq 0$ such that
\begin{equation}
y'_1 = y_1 + t, \quad y'_2 = y_2 + t, \quad  y'_3 = y_3 - t, \quad  y'_4 = y_4 - t\,,
\end{equation}
then the restricted toolbox of effects does not allow us to distinguish the two states, hence $\omega$ and $\omega'$ now represent the \textit{same} state in the restricted model.  

\begin{figure}
\centering
\includegraphics[scale=.7]{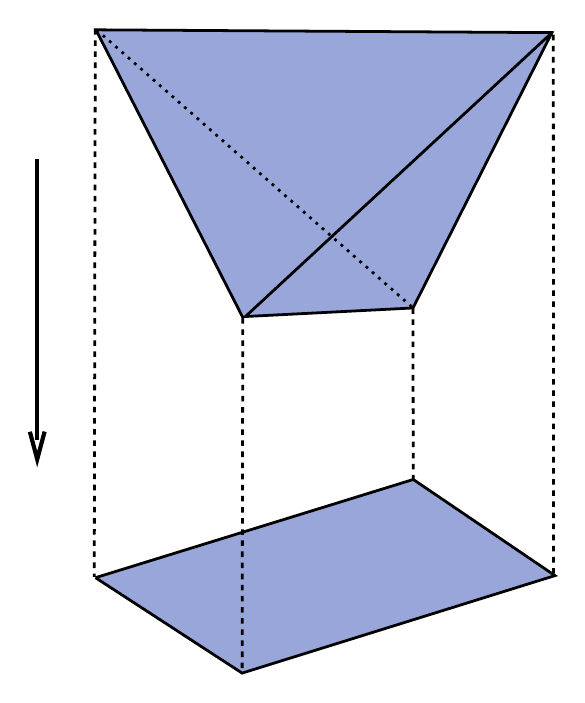}
\caption{Construction of the two-dimensional gbit state space by projecting a three-dimensional classical state space (adapted from \cite{holevo82}).}
\label{fig:projection}
\end{figure}

The extended operational equivalence allows us to choose one of the components, e.g. to set $y_4=0$. This means that the four-dimensional parameter space is projected onto a three-dimensional subspace, and the embedded three-dimensional tetrahedron is again projected to a two-dimensional convex object. This projected state space turns out to have the form of a square, as shown schematically in Fig.~\ref{fig:projection}. As we have seen before, this is just the state space of a (non-classical) gbit. Therefore, the restriction~(\ref{restriction}) leads effectively to a non-classical behavior. In fact, Holevo showed that such a construction is possible for any probabilistic theory including quantum theory\footnote{Note that quantum theory has infinitely many extremal states. The construction therefore requires an infinitely dimensional classical system.}.

To summarize, any non-classical GPT can be extended to a higher-dimensional classical theory by including additional effects. Conversely, non-classical theories can be deduced from a classical one by imposing appropriate restrictions on the available effects. The restrictions allow us to project the classical state space to a non-classical one in lower dimensions, inheriting phenomena like uncertainty relations and non-unique decompositions of mixed states. 

Obviously this seems to question the fundamental necessity of non-classical systems. How do we know that all the unusual phenomena in quantum theory do not only result from restrictions in some higher-dimensional space and thus can be explained in classical terms once we extend our theory? However, at this point we should keep in mind that so far we considered only single (non-composite) systems. As we will see in the following section, \textit{multipartite} non-classical systems cannot be described in terms of restricted classical systems. Thus, it would be misleading to conclude that non-classicality only results from restrictions imposed on an underlying higher-dimensional classical system. In fact, the analysis of multipartite systems will allow us to clearly distinguish classical and genuinely non-classical physical systems. 

\section{Multipartite systems}
\label{sec:jointsystems}

Multipartite systems may be thought of as consisting of several subsystems in which the same type of theory applies. Since such a composed system in itself can be viewed as a single system simply by ignoring its subsystems structure, the consistency conditions discussed in the previous sections obviously apply to multipartite systems as well. However, it turns out that additional consistency conditions arise from the fact that the theory has to be compatible with the partition into given subsystems. In fact, as we will see below, the structure of the subsystems determines a smallest and largest set of joint states and effects that are compatible with the given partition. The actual set of joint elements can be chosen freely within these constraints. This means that a GPT is not yet fully specified by defining states, effects, and the probability table of a single system, instead it is also required to specify how individual systems can be combined to more complex composite systems. 

We will see that, mathematically, the rules for the combination of subsystems are incorporated by defining suitable \textit{tensor products} for the sets of states and effects. These tensor products should not be confused with the (uniquely defined) Cartesian tensor product of the linear spaces $V$ and $V^*$, it rather defines the range of physical objects which are embedded in these spaces.

\subsection{Separability and the minimal tensor product}

In the simplest case, the composite system consists of several independently prepared components. As these subsystems are statistically independent, the joint state describing the overall situation is given by a product state. 

As an example let us consider two subsystems $A$ and $B$ which are in the states $\omega^A$ and $ \omega^B$, respectively. If these systems are completely independent, their joint state is given by a product state, denoted as $\omega^{AB} = \omega^A \otimes \omega^B$. Similarly, the effects of the two subsystems can be combined in product effects $e^{AB} = e^A \otimes e^B$, describing statistically independent measurements on both sides. In this situation the joint measurement probabilities factorize, i.e.
\begin{equation}
\fl\qquad
e^{AB}(\omega^{AB}) \;=\; p(e^A \otimes e^B|\omega^A \otimes \omega^B) \;=\; p(e^A|\omega^A) \, p(e^B|\omega^B) \;=\; e^A(\omega^A) \, e^B(\omega^B).
\end{equation}
As a next step, we include classical correlations by randomly choosing preparation procedures and measurement apparatuses in a correlated manner. Formally this can be done by probabilistically mixing the product elements defined above. For example, classically correlated states may be incorporated by including probabilistic linear combinations of the form $\omega^{AB}=\sum_{ij} \lambda_{ij} \, \omega_i^A \otimes \omega_j^B$ with positive coefficients $\lambda_{ij}>0$. Similarly, one can introduce classically correlated effects. 

In the GPT framework the mathematical operation, which yields the set of product elements and their probabilistic mixtures, is the so-called the \emph{minimal tensor product}:\\
\begin{eqnarray}
\fl\qquad
 \label{eq:mintensorproduct}
 V^{A}_+ \tensormin V^{B}_+ &:= \{\omega^{AB} = \sum_{ij}  \lambda_{ij}\, \omega_i^A \otimes \omega_j^B \ | \ \omega^A \in V^{A}_+, \omega^B \in V^{B}_+, \lambda_{ij} \geq 0\} \\
\fl\qquad
  V^{A*}_+ \tensormin V^{B*}_+  &:= \{e^{AB} \ = \sum_{ij} \mu_{ij} \, e_i^A \otimes e_j^B \ \ | \ e^A \in V^{A*}_+, e^B \in V^{B*}_+, \mu_{ij} \geq 0\}\,.
 \end{eqnarray}
Elements in the minimal tensor product are called \emph{separable} with respect to the partition. 

The extremal states in the joint state space $V^{A}_+ \tensormin V^{B}_+ $ are given by the product of extremal subsystem states. Note that the joint states in this space are not necessarily normalized. Normalized separable joint states can be obtained by forming products of normalized single states or mixtures of them. As a criterion for normalization, the joint unit measure $u^{AB} = u^A \otimes u^B$ is the unique joint effect that gives $u^{AB}(\omega^{AB})=1$ on all normalized joint states. 

If we apply a joint effect to a joint state, the corresponding measurement statistics is determined by the weighted sum of factorizing probabilities:
\begin{eqnarray}
 \nonumber
 e^{AB}(\omega^{AB})=p (e^{AB}|\omega^{AB}) &= \left[\sum_{i,j} \mu_{ij} \, e^A_i \otimes e^B_j \right] \left( \sum_{kl} \lambda_{kl} \, \omega^A_k \otimes \omega^B_l \right)\\ \label{eq:jointprob}
 {} &= \sum_{ijkl} \lambda_{ij} \, \mu_{kl} \, e^A_i(\omega^A_k) \, e^B_j(\omega^B_l).
\end{eqnarray}
As the number of combinations and the number of coefficients $\lambda_{ij},\mu_{kl}$ coincide, it is clear that the measurement statistics obtained from such product effects is sufficient to identify a joint state uniquely. This means that the whole information of classically correlated elements in the minimal tensor product can be extracted by coordinated \textit{local} operations carried out in each of the subsystems. 

\subsection{Entanglement in GPTs}

The minimal tensor product defines the sets $ V^{A}_+ \tensormin V^{B}_+$ and $ V^{A*}_+ \tensormin V^{B*}_+$ of classically correlated states and effects which can be seen as subsets of certain vector spaces. For \textit{classical} systems one can show that the minimal tensor product already includes all joint elements that are consistent with the division into classical subsystems \cite{nobroadcasting}. However, in non-classical theories there are generally additional vectors representing elements which are non-separable but nevertheless consistent with the subsystem structure and fully identifiable by local operations and classical communication (LOCC). Such states are called \textit{entangled}. As it is well known, entangled states do exist in standard quantum theory.

In the GPT framework both separable and entangled elements can be represented as vectors in the direct product spaces $V^{AB}=V^A \otimes V^B$ and $V^{AB*}=V^{A*} \otimes V^{B*}$. For separable, classically correlated elements this was directly inherited from classical probability theory. The tensor structure for entangled elements is based on the following additional assumptions\cite{barrett07}: i)\textit{local tomography}, which means that coordinated local operations suffice to identify a joint element, ii) \textit{no-signaling}, stating that local operations in one part of the system have no effect on the local measurement statistics in other parts. As elements of the direct product spaces, we can represent joint elements as $n \times m$ matrices, where $n = \dim V^A = \dim V^{A*}$ and  $m = \dim V^B = \dim V^{B*}$ are the dimensions of the subsystems.

Separable and entangled elements decompose into product elements in a different way. By definition, entangled elements are are not included in the minimal tensor product, meaning that they cannot be written as probabilistic mixtures of product elements. Of course they can still be decomposed into a linear combination of product elements, but such a linear decomposition would inevitably include negative coefficients. 

As an example from quantum mechanics let us consider a fully entangled two-qubit Bell state 
\begin{equation}
\label{BellState}
\ket{\psi^+}=\frac{1}{\sqrt{2}}(\ket{00}+\ket{11})\,.
\end{equation}
Choosing for each qubit the normalized extremal Bloch states 
\begin{equation}
\{\omega_1,\omega_2,\omega_3,\omega_4\}=\{\frac{\id}{2}, \frac{\id+\sigma^x}2, \frac{\id+\sigma^y}2, \frac{\id+\sigma^z}2\}\,,
\end{equation}
where $\sigma^{x,y,z}$ are Pauli matrices, a straight-forward calculation shows that the pure Bell state $\omega=\ket{\psi^+}\bra{\psi^+}$ can be decomposed into a linear combination
\begin{equation}
\omega = 2\omega_{11}-\omega_{12}+\omega_{13}-\omega_{14}-\omega_{21}+\omega_{22}+\omega_{31}-\omega_{33}-\omega_{41}+\omega_{44}\,
\end{equation}
of the product states $\omega_{ij}=\omega_i\otimes\omega_j$, which obviously contains negative coefficients.

\subsection{Entanglement as a genuinely non-classical phenomenon}

\begin{figure}
\centering
\includegraphics[width=110mm]{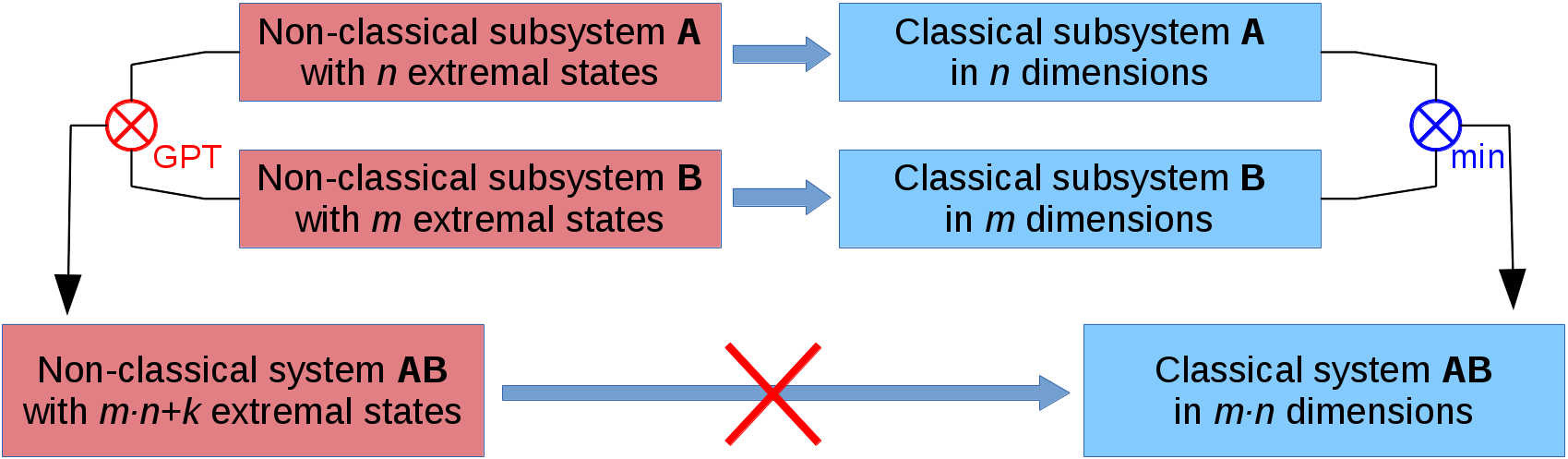}
\caption{Impossibility to explain entanglement in classical terms. Two non-classical subsystems \textbf{A},\textbf{B} containing $n$ and $m$ extremal states are joined to a single non-classical system by means of a nontrivial tensor product $\otimes_{\rm GPT} > \otimes_{\rm min}$. The resulting non-classical systems \textbf{AB} contains $n\cdot m$ extremal product states plus $k$ additional non-separable extremal states. However, mapping the systems \textbf{A,B} first to the corresponding higher dimensional classical systems (right side) and combining them by the usual classical tensor product, the resulting $m \cdot n$-dimensional classical system would not be able to account for the additional $k$ entangled states.} 
\label{fig:tensor}
\end{figure}

The previous example of the Bell state illustrates that the phenomenon of entanglement gives rise to additional extremal joint elements in the tensor product which are not part of the minimal tensor product. As we will see in the following, the existence of such non-separable elements makes it impossible to consistently `simulate' a non-classical system by a classical theory in a higher-dimensional state space.

To see this we first note that the product of extremal elements in the subsystems gives again extremal elements in the composite system. For example, if two non-classical systems $A$ and $B$ have $n$ and $m$ pure states, then the joint system $AB$ possesses at least $nm$ pure product states (see Fig.~\ref{fig:tensor}). In addition, the joint system also possesses a certain number $k$ of non-separable extremal states, provided that the tensor product is `larger' than the minimal one.

The existence of non-separable elements is incompatible with the idea of an underlying classical system in a higher-dimensional space with appropriate restriction, as described in Sect. \ref{sec:nonclassicalbyrestriction}. The reason is that the combination of two classical systems cannot account for additional non-separable elements.\footnote{This is because there are no physical states outside the classical probability simplex.}. In other words, if we first map the subsystems to the corresponding $n$- and $m$-dimensional classical systems and combine them by means of the classical (i.e. minimal) tensor product, the resulting classical would live in a $nm$-dimensional space. However, in order to account for the entangled elements, $nm+k$ dimensions would be needed, as illustrated in the figure. In other words,  such a construction shows an inconsistent scaling behavior. 

The measurement probability $p(e^{AB}|\omega^{AB})$ of an arbitrary joint effect $e^{AB}$ applied to an arbitrary joint state $\omega^{AB}$ is still given by \eref{eq:jointprob}, but in the case of non-separable elements some of the coefficients $\lambda_{ij}$ and $\mu_{kl}$ will be negative. Since this could lead to negative probabilities $p(e^{AB}|\omega^{AB})<0$, further restrictions on the joint elements are needed to ensure positivity. 

Typically, if we include more and more entangled states, the allowed range of entangled effects becomes smaller, and vice versa. Thus, as in the case of single systems there is a trade-off between entangled states and effects \cite{enttradeoff}. In particular, if we restrict the range of effects to the minimal tensor product, we can include a certain maximal set of consistent joint states and vice versa. In the following we want to characterize this maximal set of joint states.

\subsection{Marginal states and conditional states}
\label{sec:subsystems}

Before defining the maximal set of joint states, we have to introduce the notion of \textit{marginal states}. To this end let us first consider the measurement statistics of independent local measurements applied to a joint state $\omega^{AB}$, which is given by the joint probability distribution $p(e^A_i,e^B_j|\omega^{AB}) = [e^A_i \otimes e^B_j] (\omega^{AB})$. Since the local measurements are independent, we do not have to apply the effects $e^A_i$ und $e^B_j$ at once. In particular, we could observe only the outcome of $e^A$ in part $A$, ignoring the measurement in part $B$. The probability of this outcome is given by the marginal probability
\begin{eqnarray}
 \nonumber
 \fl\quad
 p(e^A_i\,|\,\omega^{AB}) &= \sum_j p(e^A_i,e^B_j\,|\,\omega^{AB}) = \sum_j [e^A_i \otimes e^B_j] (\omega^{AB}) = e^A_i \otimes \left[ \sum_j e^B_j  \right] (\omega^{AB})  \\
 \label{eq:marginal}
 {}&=\; [e^A_i \otimes u^B] (\omega^{AB}) \;=\; e^A_i (\omega^A_{u^B})\,.
\end{eqnarray} 
In the last step of Eq. \eref{eq:marginal} we introduced the so-called \textit{marginal state} $\omega^A_{u^B}$ (analogous to the reduced density matrix from the partial trace in quantum mechanics). This state is the effective subsystem state which predicts the local measurement statistics in part $A$. Similarly, the marginal state~$\omega^B_{u^A}$ determines the measurement statistics in part $B$.

It is important to note that entangled pure states can have mixed marginal states. For example, in standard quantum mechanics the pure state $\rho^{AB} = \ket{\psi^+}\bra{\psi^+}$ in Eq.~(\ref{BellState} has a completely mixed marginal state $\rho^A = \frac12 \, (\ket{0}\bra{0} + \ket{1}\bra{1})$. Such a situation, where we have perfect knowledge about the entire system but an imperfect knowledge about its parts,  is obviously impossible in classical systems. In addition, the observation leads us to the important insight that the concept of probability in GPTs is not just a matter of incomplete subjective knowledge but rather an important part of the physical laws.

The marginal state $\omega^A_{u^B}$ reflects our knowledge about subsystem $A$ provided that potential measurements on subsystem $B$ are ignored. However,  our knowledge is of course different if a particular measurement on $B$ is carried out and the result is to us via classical communication. This increased knowledge is accounted for by the conditional probabilities
\begin{equation}
\fl\quad
 p(e^A_i|e^B_j,\omega^{AB}) = \frac{p(e^A_i,e^B_j\,|\,\omega^{AB})}{p(e^B_j|\omega^{AB})} = \frac{[e^A_i \otimes e^B_j] (\omega^{AB})}{e^B_j (\omega^B_{u^A})} = e^A_i \left( \frac{\omega^{A}_{e^B_j}}{e^B_j (\omega^B_{u^A})} \right) 
 = e^A_i \left( \tilde{\omega}^{A}_{e^B_j} \right)  \,.
 \label{eq:condprobs} 
\end{equation}
In the last steps we introduced the so-called \textit{conditional state} $\omega^{A}_{e^B_j}$ and its normalized version $\tilde{\omega}^{A}_{e^B_j}$. The conditional state $\tilde{\omega}^{A}_{e^B_j}$ is the effective state in $A$ given that the effect $e^B_j$ was observed in $B$. The marginal state introduced in Eq. (\ref{eq:marginal}) is a special conditional state, where the effect $e_j^B$ is just the unit measure in $B$.

As $\omega^{A}_{e^B_j}$ depends on the effect $e^B_j$, it can be interpreted as a linear map from effects in one part onto conditional states in the other.

\subsection{The maximal tensor product}

As outlined above, consistency conditions lead to a trade-off between the sizes of state and effect spaces. Therefore, in order to derive the maximal set of possible joint states, let us assume that the corresponding set of effects is given by the minimal tensor product. Consequently, the joint states have to satisfy two consistency conditions: 
\begin{enumerate}
\item Applied to product effects they have to give non-negative results.
\item They induce valid conditional states, that is, all conditional states have to be elements of the corresponding subsystem state space.
\end{enumerate}
Note that the second condition always implies the first one, since all factors in (\ref{eq:condprobs}) are non-negative for any valid conditional state. Conversely, in non-restricted systems the first condition implies the second one. Therefore, it suffices to consider the non-negativity condition alone. With this assumption, the maximal set of non-normalized joint states $V^A_+ \tensormax V^B_+$ for unrestricted systems is just given by the dual cone with respect to product effects:
\begin{eqnarray}
 \label{eq:tradtensormax}
 V^A_+ &\tensormax V^B_+ \;:=\; (E^A_+ \tensormin E^B_+)^* \\
 \nonumber
 {}&= \left\{ \omega^{AB} \in V^A \otimes V^B \middle| e^A \otimes e^B (\omega^{AB}) \geq 0 \quad\forall e^A \in E^A, e^B \in E^B\right\}\,.
\end{eqnarray}
In other words, the maximal tensor product $V^A_+ \tensormax V^B_+$ is simply the set of all joint states which give nonnegative results if we apply effects from the minimal tensor product.

\begin{figure}
\centering
\includegraphics[width=50mm]{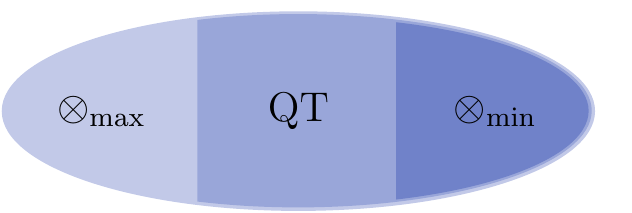}
\caption{The tensor product of standard quantum theory (QT) lies strictly between the minimal and the maximal tensor product. All tensor products share the extremal states of the minimal tensor product. Quantum theory and the maximal tensor product append additional extremal states.}
\label{fig:QTtensor}
\end{figure}

\subsection{Maximal tensor product for systems with restricted effects}

For subsystems with restricted effect spaces $E^A$, $E^B$ the situation is more complicated. In this case the spaces $V^A$, $V^B$ contain joint elements that give probability-valued results, but yield invalid elements as conditional states. Consequently, the construction in (\ref{eq:tradtensormax}) is no longer suitable. 

To circumvent this problem a generalized maximal tensor product denoted by $\gtensormax$ was proposed in \cite{norestriction}. The idea is to construct a maximal extension in only one direction, say from subsystem $A$ to subsystem $B$, then to repeat this construction and opposite direction, and finally to define the generalized tensor product as the intersection of both results.

\begin{figure}
 \centering
 \includegraphics[width=120mm]{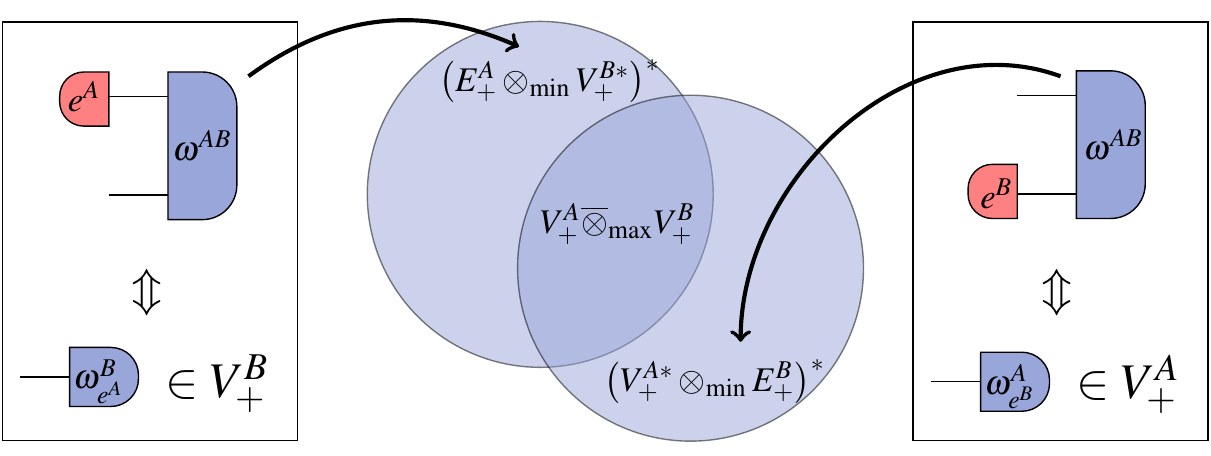}
 \caption{Illustration of the generalized maximal tensor product (originally published in \cite{norestriction})}
 \label{fig:gentensormaxconstruction}
\end{figure}

Let us first consider the $A \to B$ direction. Recall that the maximal tensor product should give all linear maps $\omega_{e^A}^B$ from effects in $A$ to valid non-normalized states in the cone $V^B_+$. Consequently, we get can maximally extend the states in $A$ and take the full dual cone $V^{B*}_+$ for effects in $B$ without affecting the valid linear maps in this direction. In this way, we obtain an intermediate unrestricted theory to which we can apply the previous construction in Eq. (\ref{eq:tradtensormax}). The same can be done in opposite direction by exchanging $A \leftrightarrow B$. The generalized maximal tensor product $V^A_+ \gtensormax V^B_+$ is given by the set of linear maps that are valid in both directions, i.e. it is given by the intersection
\begin{equation}
 \label{eq:gentensormax}
 V^A_+ \gtensormax V^B_+ \;:=\; \left( E^A_+ \tensormin V^{B*}_+ \right)^* \cap \left( V^{A*}_+ \tensormin E^B_+ \right)^*\,.
\end{equation} 
The construction is illustrated in figure~\ref{fig:gentensormaxconstruction}. Note that \eref{eq:gentensormax} reduces to \eref{eq:tradtensormax} in the non-restricted case, as $E^A_+ = V^{A*}_+$ and $E^B_+ = V^{B*}_+$. Moreover, it was recently shown that it reduces to the traditional maximal tensor product of extended effects as soon as one of the subsystems is non-restricted \cite{QPL13}.

Recall that the minimal and maximal tensor products define only the boundary cases for joint systems and that there is a broad range of theories in between. Standard quantum mechanics is special in so far as it allows the same degree of entanglement for both states and effects, i.e. states and effects play a symmetric role in this theory. That is the tensor product of quantum theory lies strcitly between the minimal and the maximal tensor product. Theories at the upper boundary defined by the maximal tensor product admit a higher degree of entanglement, but only either for states or for effects, while elements of the other type have to be separable. 
%

A popular toy theory that uses the maximal tensor product is called \textit{boxworld}~\cite{gross10}. It is defined in terms of gbits or higher-dimensional variants combined with the maximal tensor product.

\section{Realism versus locality: The meaning of non-local correlations}
\label{sec:nonlocalcorrelations}

As shown above entanglement is a strictly non-classical feature of GPTs. However, a detailed knowledge of the subsystems is required in order to identify a joint element as an entangled one. Therefore, it would be desirable to formulate alternative device-independent criteria for entanglement which do not rely on the particular type of theory in the subsystems. 

The most important approach in this direction is the analysis of nonlocal correlations. Already in the early days of quantum mechanics, it was pointed out in the context of the famous EPR gedankenexperiment\cite{EPR} that the existence such non-classical correlations are in conflict with at least one of the following assumptions:

\begin{description}
\item[Realism:] A theory obeys \em realism \em if measurement outcomes can be interpreted as revealing a property of the system that exists independent of
the measurement.
\item[Locality:] A physical theory obeys \em locality \em if the measurement on one part of a joint system does not influence measurements on other (spatially separated) parts.
\end{description}
Classical systems satisfy local realism. However, Bell already showed in 1964 that quantum theory can generate correlations that violate at least one of these assumptions~\cite{bell64}. Interestingly, stronger entanglement does not always lead to stronger violations of a Bell inequality. In fact, it has been shown that in some setups even an inverse relationship is possible \cite{vallone13}.

A particularly simple and popular setup that detects non-local correlations was suggested by Clauser, Horne, Shimony and Holt (CHSH) in 1969 \cite{CHSH}. The CHSH setup, which was originally designed for quantum-mechanical systems, fits naturally into the GPT framework. It consists of two (spatially separated) parties $A$ and $B$ that share a bipartite state~$\omega^{AB}$. Each of the parties chooses between two binary measurements $M^A_x = \{e^A_{x,0},e^A_{x,1}\}$ and $M^B_y = \{e^B_{y,0},e^B_{y,1}\}$ indexed by $x, y \in \{0,1\}$. For each choice of $x,y$ we get a probability distribution
\begin{equation}
 p(a,b \, | \, x,y) = \bigl[e^A_{x,a} \otimes e^B_{y,b}\bigr] (\omega^{AB})\,,
\end{equation}
where $a,b\in\{0,1\}$. The probability distribution generated by a \textit{local realistic} theory can be shown to satisfy the \textit{CHSH inequality}
\begin{equation}
S^{\mathrm{LH}} = |C_{0,0} + C_{0,1} + C_{1,0} - C_{1,1}| \leq 2
\end{equation}
with the correlators
\begin{equation}
C_{x,y} = \sum_{a,b \in \{0,1\}} (-1)^{a \oplus b} \, p(a,b \, | \, x,y),
\end{equation}
where $\oplus$ denotes a logical XOR. 

In a classical probability theory we have $C_{0,0} = C_{0,1} = C_{1,0} = 1$, implying that the fourth correlator is given by $C_{1,1} = 1$ so that the inequality holds. Quantum theory, however, can violate this inequality as confirmed experimentally in~\cite{aspect}. The theoretical maximum of the CHSH value that can be reached in quantum theory is given by Tsirelson's bound $S^{QT}_\mathrm{max} = 2 \, \sqrt{2}$ \cite{tsirelson80}. 

Non-classical GPTs can also violate the CHSH inequality. In fact, these violations can even exceed Tsirelson's bound for quantum-mechanical systems. A frequently studied example is the maximal tensor product of gbits which exhibits so-called PR-box correlations \cite{PRbox}, violating the CHSH inequality up to its algebraic maximum $S^{PR} = 4$.

Returning to the question of what distinguishes quantum mechanics as the fundamental theory of nature, it is therefore not sufficient to explain the existence of non-classical correlations, one also has to give reasonable arguments why these correlations are not stronger.

\section{Discussion: Special multipartite features in quantum theory}
\label{sec:QTjointfeatures}

As we have seen the phenomena of nonlocality and entanglement are a hallmark of quantum theory but they also exist in many other toy theories. However, in the context of multipartite systems quantum mechanics exhibits various characteristic features which generically do not exist in other GPTs. In this section we are going to review some of these characteristic multipartite quantum features. 

\subsection{Quantum features inherited from subsystems: }
%
Surprisingly, many of these characteristic features are not linked to the structure of the tensor product, rather they are consistently inherited from single systems. In particular the one-to-one correspondence between state spaces and their information capacity carries over to joint systems. For example, the information capacity of a joint system is simply the product of the single systems information capacities \cite{hardy01}. 

Since the state spaces are uniquely determined by the information capacity, a joint system consisting of two qubits has the same state space than a single quantum system with information capacity of four. In the case of quantum mechanics the difference between single and joint systems is not reflected in different state space structures but only in a different interpretation of the measurements and states. As composite quantum systems have the same state space structure as their building blocks, multipartite quantum systems inherit all the features from single systems, including e.g. reversible continuous transformations between pure states (transitivity), strong self-duality, non-restricted measurements/states, sharpness and homogeneity. This allows us to interpret the qubit as a fundamental information unit from which any quantum system can be built~\cite{masanes13}. 

In quantum mechanics the equivalence of systems with equal information capacity also manifests itself in the associativity of the tensor product. This means that equal components of a multipartite system can be swapped without changing the state space (compound permutability) \cite{hardy11}. As illustrated in Fig.~\ref{fig:QTtensor}, this tensor product lies strictly between the minimal and maximal tensor product, such that potential entanglement in states and effects is perfectly balanced. 

The inheritance of such features to joint system in standard quantum theory is quite exceptional, as can be seen when trying to construct something similar for other GPTs. For example, the extremal states of a single gbit can be reversibly transformed into each other and there is an isomorphism between states and effects. However, as shown in~\cite{gross10} the joint states cannot be reversibly transformed. A tensor product which inherits an isomorphism between joint states and effects can be constructed, but it was shown to treat equal subsystems differently \cite{generalbox}.

Given local quantum systems it has been shown that the ordinary quantum tensor product is the only one that preserves transitivity \cite{qreversibility}. Another work explores the opposite direction \cite{ent3D}. The authors assume transitivity on joint states and local systems with state spaces bounded by hyperspheres, which is a generalization of the three-dimensional Bloch ball of qubits known as a \textit{hyperbit} \cite{hyperbit}. It was shown that entangled states in such a scenario are only possible for dimension three \cite{ent3D}, which was used in \cite{spatial3D} to explain why we are living in a three-dimensional world.  

\subsection{Genuine multipartite quantum features: }
%
Beyond the features inherited from single systems there are also genuine multipartite features that are characteristic for quantum theory. Recall that for entangled states the marginal state is mixed even though the joint state might be extremal. Quantum theory allows also the opposite, namely, any mixed state can be regarded as the marginal of a pure extremal state -- a process called \textit{purification}~\cite{GNS1, GNS2, chiribella11}. As a consequence any stochastic mapping from one mixed state to another can be realized as a reversible unitary transformation in a higher-dimensional state space without information loss~\cite{stinespring}. 

Since there is a continuum of mixtures, the possibility of purification requires a continuum of pure entangled states. There is also an isomorphism between the transformations of single systems and bipartite joint states \cite{choi72, jamiolkowski72}. This was recently used to generalize Bayesian inference to quantum states \cite{quantuminference}. This isomorphism can be further decomposed into two components. On the one hand, using the self-duality of quantum systems, states are converted to corresponding effects given by the same operator. On the other hand the transformation itself can be realized via \textit{steering}, i.e. the ability to obtain any state as the conditional state of a joint system~\cite{Schroedinger35, steering}. Steering is a prerequisite of more advanced multipartite quantum features, like quantum teleportation \cite{qteleportation, teleportation} and entanglement swapping~\cite{repeaters}. 

As pointed out before, nonlocal correlations are a central feature of quantum theory. Several articles have examined the relation between entanglement and non-local correlations in quantum theory. As quantum theory balances entanglement in states and effects, extending the joint state space to the maximal tensor product would potentially allow for new correlations. While this is not the case for bipartite systems~\cite{barnum10}, it was found that the maximal tensor product of multipartite systems with more than two subsystems can indeed generate new correlations that are not possible in the standard tensor product \cite{acin10}. 

Not only entanglement but also the local structure of the subsystems influences nonlocal correlations. For example, it was shown in \cite{locallimits} that joint states that resemble the maximally entangled states in quantum theory exist in the maximal tensor product for a class of toy theories with regular polygon shaped subsystems. It turned out that the possible nonlocal correlations strongly depend on the subsystems' structure. For bipartite systems consisting of two identical polygons with an odd number of vertices the possible correlations were found to be strictly weaker as in the quantum case. Polygons with an odd number of vertices, however, yield correlations stronger or equal than quantum correlations. A general connection between nonlocality and uncertainty relations of subsystems has been found in Ref.~\cite{oppenheim10}. 

\begin{figure}
\centering
\includegraphics[scale=1]{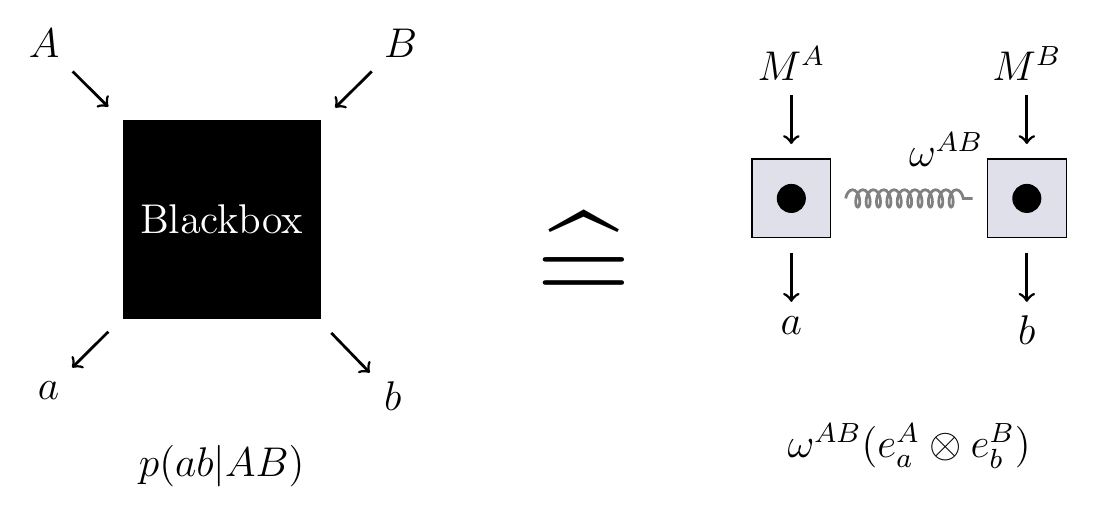}
\caption{Nonlocal boxes as a device-independent model of nonlocal correlations abstracting from specific measurements and states.}
\label{fig:iobox}
\end{figure}

The device-independent view on nonlocal correlations studies correlations independent of specific joint states and local measurements, simply by considering the probabilities of input-output combinations for a given choice of measurement as input and the outcomes as outputs (see Fig.~\ref{fig:iobox}). These so-called nonlocal boxes allow us to compare general no-signaling correlations to those possible in quantum theory without referring to specific theories. For the CHSH setup the no-signaling correlations form an eight-dimensional polytope with the classical correlations and the PR box with maximal nonlocality as extremal points~\cite{NLinforesource, Scarani09}. The quantum correlations form a convex subset with infinitely many extremal points. This set can be determined by a infinite hierarchy of semi-definite programs, whereas an analytical upper bound known as $Q_1$ has been derived from the first order \cite{Q1-1,Q1-2}. It was shown that any theory that is able to recover classical physics in the macroscopic limit has correlations limited by this bound \cite{ML}. Also $Q_1$ obeys information causality \cite{IC, Q1IC}. That is given the nonlocal resource and $m$ bits of classical communication a party on one side can learn at most $m$ bits about the system on the other side. Note that this is a generalization of no-signaling that refers to the situation with $m=0$. Different to quantum correlations general no-signaling correlations can violate information causality up to the extreme cases of PR boxes that can evaluate any global function that depends on both local inputs from only one classical bit of communication (trivial communication complexity) \cite{vandam05}. Interestingly, for some nonlocal boxes given multiple copies allows to distill PR boxes by using only classical processing at each of the local parts individually \cite{distill}, whereas the quantum correlations are closed under such operations \cite{closedsets}.

In conclusion quantum theory a lot of unique characteristic physical features. The framework of Generalized Probabilistic Theories presented in this review paper played a crucial role to identify many of those.

\ack
We thank Markus M\"uller for discussions and in particular for pointing out the Holevo construction of a gbit by a higher dimensional restricted classical system. PJ is supported by the German Research Foundation (DFG). 

\appendix
\section{Partial order of effects}
\label{app:partialorder}

Given two effects $e_i$, $e_j$ one is said to be dominated by the other iff it occurs with lower probability for any state:
\begin{equation}
\label{eq:partial1}
 e_i \leq e_j \Leftrightarrow e_i(\omega) \leq e_j (\omega) \forall \omega \in \Omega
\end{equation}
Note that there are effects that cannot be compared in such a way. There might be states that give higher probabilities for $e_i$, while other states give higher probabilities on $e_j$. Therefore this is called a partial order.

A partial order on the elements of a vector space can be induced by a convex cone. The partial order of effects is based on the dual cone $V^*_+$ with
\begin{equation}
\label{eq:partial2}
 e_i \leq e_j \Leftrightarrow e_i \in e_j - V^*_+.
\end{equation} 
To see that this is equivalent to \eref{eq:partial1} recall that the dual cone $V^*_+$ is given by all elements in $V^*$ with non-negative results on the state space elements $\omega \in \Omega$. Consequently, subtracting one of these elements from an effect cannot result in a bigger value for any state.

Note that the dual cone depends on the state space $\Omega$. An extension of a model with new states might therefore also affect the partial order.  

\section*{References}


\end{document}